\documentclass[bibyear]{aa}
\usepackage{graphicx}
\usepackage{txfonts}
\begin{document} 

\title{The contribution of faint AGNs to the ionizing background at
$z\sim 4$\thanks{Based on
observations made at the Large Binocular Telescope
(LBT) at Mt. Graham (Arizona, USA). Based on observations
collected at the European Organisation for Astronomical
Research in the Southern Hemisphere under ESO programme 098.A-0862.
This paper includes data gathered with the 6.5 meter Magellan Telescopes
located at Las Campanas Observatory, Chile.}}

\author{A. Grazian\inst{1}
\and E. Giallongo\inst{1}
\and K. Boutsia\inst{2}
\and S. Cristiani\inst{3}
\and E. Vanzella\inst{4}
\and C. Scarlata\inst{5}
\and P. Santini\inst{1}
\and L. Pentericci\inst{1}
\and E. Merlin\inst{1}
\and N. Menci\inst{1}
\and F. Fontanot\inst{3}
\and A. Fontana\inst{1}
\and F. Fiore\inst{1}
\and F. Civano\inst{6,7}
\and M. Castellano\inst{1}
\and M. Brusa\inst{8}
\and A. Bonchi\inst{9,1}
\and R. Carini\inst{1}
\and F. Cusano\inst{4}
\and M. Faccini\inst{1}
\and B. Garilli\inst{10}
\and A. Marchetti\inst{11}
\and A. Rossi\inst{12}
\and R. Speziali\inst{1}
}

\offprints{A. Grazian, \email{andrea.grazian@oa-roma.inaf.it}}

\institute{INAF--Osservatorio Astronomico di Roma, Via Frascati 33,
I-00078, Monte Porzio Catone, Italy
\and
Las Campanas Observatory, Carnegie Observatories, Colina El Pino Casilla 601,
La Serena, Chile
\and
INAF--Osservatorio Astronomico di Trieste, Via G.B. Tiepolo 11
I-34143, Trieste, Italy
\and
INAF--Osservatorio Astronomico di Bologna, via P. Gobetti 93/3, I-40129,
Bologna, Italy
\and
Minnesota Institute for Astrophysics, University of Minnesota,
116 Church Street SE, Minneapolis, MN 55455, USA
\and
Yale Center for Astronomy and Astrophysics, 260 Whitney
Avenue, New Haven, CT 06520, USA
\and
Harvard-Smithsonian Center for Astrophysics, 60 Garden
Street, Cambridge, MA 02138, USA
\and
Dipartimento di Fisica e Astronomia, Universit\`a di Bologna,
viale Berti Pichat 6/2, I-40127 Bologna, Italy
\and
ASI Space Science Data Center, Via del Politecnico snc,
I-00133 Rome, Italy
\and
INAF--Istituto di Astrofisica Spaziale e Fisica Cosmica di Milano,
via Bassini 15, I-20133, Milano, Italy
\and
INAF--Osservatorio Astronomico di Padova, Vicolo Osservatorio 5, I-35122,
Padova, Italy
\and
INAF--Istituto di Astrofisica Spaziale e Fisica Cosmica di Bologna,
via P. Gobetti 101, I-40129, Bologna, Italy
}

\date{Received Month day, year; accepted Month day, year}

\authorrunning{Grazian et al.}
\titlerunning{The ionizing radiation of AGNs at $z>4$}
 
  \abstract
{
Finding the
sources responsible for the hydrogen reionization is one of
the most pressing issues in observational cosmology. Bright QSOs are known to
ionize their surrounding neighborhood, but they are too few to ensure
the required HI ionizing background. A significant contribution by
faint AGNs, however, could solve the problem, as recently advocated on the
basis of a relatively large space density of faint active nuclei at $z>4$.
}
{
This work is part of a long term project aimed at measuring
the Lyman Continuum escape fraction for a large sample of AGNs at $z\sim 4$
down to an absolute magnitude of $M_{1450}\sim -23$.
We have carried out an exploratory spectroscopic program to measure the HI
ionizing emission of 16 faint AGNs spanning a broad $U-I$ color interval,
with $I\sim 21-23$ and
$3.6<z<4.2$. These AGNs are three magnitudes fainter than the typical
SDSS QSOs ($M_{1450}\lesssim -26$) which are known to ionize their surrounding
IGM at $z\gtrsim 4$.
}
{
We acquired deep spectra of these faint AGNs with
spectrographs available at the VLT, LBT, and Magellan telescopes, i.e.
FORS2, MODS1-2, and LDSS3, respectively. The emission in the Lyman Continuum
region, i.e. close to 900 {\AA} rest frame, has been detected with S/N ratio of
$\sim 10-120$ for all the 16 AGNs. The flux ratio between the 900 {\AA} rest
frame region and 930 {\AA} provides a robust estimate of the escape fraction
of HI ionizing photons.
}
{
We have found that the Lyman Continuum escape fraction
is between 44 and 100\% for all the observed
faint AGNs, with a mean value of 74\% at $3.6<z<4.2$ and
$-25.1\lesssim M_{1450}\lesssim -23.3$, in agreement with the value found
in the literature for
much brighter QSOs ($M_{1450}\lesssim -26$) at the same redshifts.
The Lyman Continuum escape fraction of our faint AGNs
does not show any dependence on the absolute luminosities or on the observed
$U-I$ colors of the objects. Assuming that the Lyman Continuum escape
fraction remains close to
$\sim 75\%$ down to $M_{1450}\sim -18$, we find that the AGN population
can provide between 16 and 73\% (depending on the adopted luminosity
function) of the whole ionizing UV background
at $z\sim 4$, measured through the Lyman forest.
This contribution increases to 25-100\% if other determinations of the
ionizing UV background are adopted from the recent literature.
}
{
Extrapolating these results to $z\sim 5-7$, there
are possible indications that bright QSOs and faint AGNs
can provide a significant
contribution to the Reionization of the Universe, if their space
density is high at $M_{1450}\sim -23$.
}


   \maketitle
%
%-------------------------------------------------------------------

\section{Introduction}

One of the most pressing questions in observational cosmology is
related to the reionization of neutral hydrogen (HI) in the
Universe. This fundamental event marks the end of the so-called
Dark Ages and is located in the redshift interval $z=6.0-8.5$. The lower
limit is derived from observations of the Gunn-Peterson effect in
luminous $z>6$ QSO spectra (Fan et al. 2006), while the most recent
upper limit, $z<8.5$, comes from measurements of the Thomson optical
depth $\tau_e=0.055\pm 0.009$ in the CMB polarization map by Planck
(Planck Collaboration 2016). While we now have a precise timing of the
reionization process, we are still looking for the sources providing
the bulk of the HI ionizing photons. Obvious candidates include
high-redshift star-forming galaxies (SFGs) and/or Active Galactic
Nuclei (AGNs).

High-z SFGs have been advocated as the most natural way of explaining
the reionization of the Universe (Robertson et al. 2015; Finkelstein
et al. 2015; Schmidt et al. 2016; Parsa et al. 2018). The two
critical ingredients in modeling reionization are the relative escape fraction
of HI ionizing photons $f_{esc,rel}$ (and its luminosity and redshift
dependence) and the number density of faint galaxies which can be
measured by a precise evaluation of the faint-end slope of the UV
luminosity function at high-z. Finkelstein et al. (2012, 2015) and
Bouwens et al. (2016) show that an $f_{esc,rel}$ of $\ge 10-20\%$\footnote{The
exact value slightly depends also on the adopted
clumping factor $C_{HII}$ and on the Lyman continuum
photon production efficiency $\xi_{ion}$.} must be assumed for {\em all}
the galaxies down to
$M_{1500}=-13$ in order to keep the Universe ionized at $3\le z\le 7$, and
that the luminosity function should be steeper than $\alpha\sim -2$ in
order to have a large number of faint sources. This latter assumption
has been recently confirmed by the steep luminosity function
found by Livermore et
al. (2017) and Ishigaki et al. (2017) in the HST Frontier Fields, down
to $M_{1500}=-12.5$ at $z=6$ and in the MUSE Hubble Ultra Deep Field by
Drake et al. (2017), but see Bouwens et al. (2017) and Kawamata et al.
(2017) for different results.

The search for HI ionizing photons escaping from SFGs has
not been very successful. At $z<2$ all surveys
appear to favour low $f_{esc,rel}$ from relatively bright galaxies, and
recent limits on $f_{esc,rel}$ are below 1\% (Grimes et al. 2009; Cowie et
al. 2009; Bridge et al. 2010). At fainter magnitudes, Rutkowski et
al. (2016) found that $f_{esc,rel}<5.6\%$ for $M_{1500}<-15$ ($L>0.01L^*$)
galaxies at $z\sim 1$, concluding that these SFGs contribute less than
50\% of the ionizing background. Few exceptions have been found
at $z<0.4$, e.g. Izotov et al. (2016a,b) found five galaxies with
$f_{esc,rel}=9-34\%$, while Leitherer et al. (2016) found two
galaxies\footnote{One of these is Tol 1247-232, probably a low-luminosity
AGN, see Kaaret et al. (2017).} with $f_{esc,rel}=20.8-21.6\%$.

At higher redshift ($z>2$), there are contrasting results: Mostardi et
al. (2013) found $f_{esc,rel}$ of $\sim 5-8$\% for Lyman Break galaxies
($f_{esc,rel}\sim 18-49\%$ for Lyman-$\alpha$ emittes), but their samples
could be partly contaminated by foreground objects due to the lack of
high spatial resolution imaging from HST (Vanzella et al. 2010; Siana
et al. 2015). Recently, three Lyman Continuum (LyC)
emitters have been confirmed within the
SFG population at $z\sim 3$ (Vanzella et al. 2016; Shapley et al. 2016;
Bian et al. 2017). Other teams, using both spectroscopy and very deep broad
or narrow band imaging from ground based telescopes and HST, give only
upper limits in the range $<2-15$\% (Grazian et al. 2016, 2017; Guaita et
al. 2016; Vasei
et al. 2016; Marchi et al. 2017; Japelj et al. 2017; Rutkowski et
al. 2017). These limits cast serious doubts on any redshift evolution
of $f_{esc}$, if the observed trend at $z\lesssim 3$ is extrapolated at higher-z.
The low $f_{esc}$ suggests that we may have a
problem in keeping the Universe ionized with just SFGs
(Fontanot et al. 2012; Grazian et al. 2017; Madau 2017).

There is therefore room for a significant contribution made by AGNs at
$z>3$. It is well known (see e.g. Prochaska et al. 2009; Worseck et
al. 2014, Cristiani et al. 2016) that bright QSOs
($M_{1450}\le -26$ at $z\ge 3$) are
efficient producers of HI ionizing photons, and they can ionize large
bubbles of HI even at distances up to several Mpc out to $z\sim 6$.
However, their space density is too low at high-z to provide the
cosmic photo-ionization rate required to keep the IGM ionized at $z>3$
(Fan et al. 2006; Cowie et al. 2009; Haardt \& Madau 2012). The bulk
of ionizing photons could then come from a population of fainter AGNs.
Interestingly, the recent observations of an early and extended period
for the HeII reionization at $z\sim 3-5$ by Worseck et al. (2016)
seem to indicate that hard ionizing photons from AGNs, producing
strongly fluctuating background on large scales, may in fact be
required to explain the observed chronology of HeII reionization.
Moreover, the observations of a constant ionizing UV background (UVB)
from z=2 to z=5 (Becker \& Bolton 2013) and
the presence of long and dark absorption troughs at $z\ge 5.5$ along
the lines of sight of bright $z=6$ QSOs (Becker et al. 2015)
are difficult to be reconciled
with a population of ionizing sources with very high space densities
and low clustering, such as ultra-faint galaxies
(Madau \& Haardt 2015; Chardin et al. 2016, 2017).
This leaves open the possibility that faint AGNs ($L\lesssim L^*$)
at $z\ge 3-5$ could be
the major contributors to the ionizing UV background.

Deep optical surveys at $z=3-5$ with complete spectroscopic
information (Glikman et al. 2011) are showing the presence of a
considerable number of faint AGNs ($L\lesssim L^*$)
producing a rather steep luminosity
function. This result has been confirmed and extended to fainter
luminosities ($L\lesssim 0.1 L^*$) by Giallongo et al. (2015)
by means of NIR (UV
rest-frame) selection of $z>4$ AGN candidates with very weak X-ray detection
in the deep CANDELS/GOODS-South field (but see Parsa et al. 2018,
Hassan et al. 2018, and Onoue et al. 2017 for different results).
The presence of a faint ionizing population of AGNs
could, if confirmed, strongly contribute to the ionizing UV
background (Madau \& Haardt 2015; Khaire et al. 2016), provided that a
significant fraction of the produced LyC photons is free to escape
from the AGN host galaxy even at faint luminosities.

Recently, the LyC $f_{esc}$ of a bright $R\le 20.15$ AGN sample
($L\gtrsim 15 L^*$, or $M_{1450}\le -26$) was estimated by Cristiani
et al. (2016). Approximately 80\% of the sample shows large LyC
emission ($f_{esc}\sim 100\%$), while $\sim$20\% are not emitting at
$\lambda<900$ {\AA} rest-frame, possibly due to the presence of a
broad absorption line (BAL) or associated absorption systems (e.g., Damped
Lyman-$\alpha$ systems, DLAs).
No trend of $f_{esc}$ with UV luminosity is detected by Cristiani
et al. (2016), though the
explored range in absolute magnitude is small, and there is not enough leverage
to draw firm conclusions at the moment.
Micheva et
al. (2017) have studied the escape fraction of a small sample of
$R\sim 21-26$ AGN (both type 1 and type 2) at $z\sim 3.1-3.8$ in the
SSA22 region through narrow band imaging. They concluded that the
contribution of these faint AGNs ($-24\lesssim M_{1450}\lesssim -22$)
is not exceeding $\sim 20\%$ of the
total ionizing budget. However, due to the limited depth of
their narrow band UV images, the broad luminosity range and the broad redshift
interval of their sample, their constraints on $f_{esc}$ for the
faint AGN population are probably not conclusive.
As an example, it is worth mentioning the
results of
Guaita et al. (2016) on faint AGNs in the CDFS region, where a large escape
fraction ($\ge 46\%$) was found for 2 intermediate luminosity
($M_{1450}\sim -22$) AGNs at $z\sim 3.3$ with deep narrow band imaging, while
less rigid upper limits have been measured for 6 AGNs at similar
redshifts and luminosities (mean $f_{esc}\le 45\%$).

In this paper we carry out a systematic survey of the Lyman
continuum escape fraction for faint AGNs ($L\sim L^*$) at
$3.6\le z\le 4.2$ through deep optical/UV spectroscopy.
This redshift interval is a good compromise between minimizing the 
IGM absorption intervening
along the line of sight and allowing observations from the ground.

This paper is organized as follows. In Sect. 2 we present the data-set,
in Sect. 3 we describe the method adopted, in Sect. 4 we show the
results for individual objects and for the overall sample as a whole,
providing an estimate of the ionizing background produced
by AGNs at $z\sim 4$. In Sect. 5 we discuss the robustness of
our results and in Sect. 6 we
provide a summary and the conclusions. Throughout the paper we adopt
the $\Lambda$-CDM concordance cosmological model ($H_0 = 70~
km/s/Mpc$, $\Omega_M=0.3$ and $\Omega_\Lambda=0.7$), consistent with recent
CMB measurements (Planck collaboration 2016). All magnitudes
are in the AB system.

%--------------------------------------------------------------------

\section{Data}

\subsection{The AGN sample}

In order to quantify the HI ionizing emission of the whole AGN
population, we carried on an exploratory spectroscopic program to measure
the LyC escape fraction of a small sample of faint galactic nuclei,
with $I\sim 21-23$ and $3.6<z<4.2$. These AGNs are three magnitudes
fainter than the typical SDSS QSOs ($M_{1450}\sim -26$) which are known to
ionize their surrounding IGM at $z>3$ (Prochaska et al. 2009;
Cristiani et al. 2016).
At present it is not known in fact whether the
escape fraction of AGN scales with their luminosity and it is not obvious
that faint AGNs have the same escape fraction
($f_{esc}\sim 75-100\%$) typical of brighter QSOs.

We selected the redshift interval $3.6\le z\le 4.2$ for the following reasons:
first, at
these redshifts the mean IGM transmission is still high ($\sim 30\%$)
compared to the large opacities found at $z\ge 6$ (e.g. Fan et
al. 2006), which prevent a direct measurement of ionizing
photons at the reionization epoch; second, at these redshifts there are a
number of relatively faint AGNs ($L\lesssim L^*$)
with known spectroscopic redshifts already available, while it is very
difficult to assemble a similar sample at $z\ge 5$; third, at $z\sim 4$
sub-$L^*$ AGNs are still bright enough to be studied in
detail with 8-10m class telescopes equipped with efficient instruments
in the near UV.

Our targets have been selected from the COSMOS (Marchesi et
al. 2016; Civano et al. 2016), NDWFS, DLS (Glikman et al. 2011), and
the SDSS3-BOSS (Dawson et al. 2013) surveys. From the parent sample of
951 AGNs with $3.6<z_{spec}<4.2$ and $21<I<23$, we have randomly identified 16
objects with $2.0\lesssim U-I\lesssim 5.0$. These limits are
approximately the minimum and maximum colors of the parent AGN sample,
indicating that this small group of 16 objects is not affected by
significant biases in their $f_{esc}$ properties, but represents an
almost uniform coverage of the color-magnitude distribution for $z\sim
4$ AGNs. At the redshifts probed by our sample, the U filter partially
covers the ionizing portion of the spectra.
However, it is worth stressing that, due to the stochasticity of the
IGM absorption and the broad filters adopted, it is not possible to
directly translate the $U-I$ color into a robust value of $f_{esc}$,
and UV spectroscopy is thus required. More precisely, the $U-I$ color
distribution is a proxy for intervening absorptions rather than an
indicator of LyC escape fraction for $z\sim 4$ AGNs. As we will show in the
following sections, their $U-I$ colors, their apparent I-band
magnitudes and their intrinsic luminosities are not biased against or in favour
of objects with peculiar properties and can thus be representative of
the whole population of faint AGNs at high-z.
The selected AGNs cover a large
interval in right ascension in order to facilitate the scheduling of
the observations and are selected both in the northern and in the
southern hemispheres in order to be targeted by many observational
facilities, e. g. with LBT, VLT, and Magellan. The limited number (16) of
selected AGNs was chosen to keep the total observing time of the
order of a normal observing program (1-2 nights).
This initial group of 16 AGNs with $M_{1450}<-23$ does not represent a
complete sample of all the type 1 and 2 active galactic nuclei
brighter than $L^*$, since the goal of these initial observations was
first to show that the program was feasible and the requested exposure
time was sufficient to estimate the LyC escape fractions of these
objects with small uncertainties.

In addition to the 16 known AGN described above, we have found a
serendipitous faint AGN at $z\sim 4.1$ (UDS10275) during a
spectroscopic pilot project with the Magellan-II Clay telescope. Since
this object is relatively bright ($I=22.27$) and the available
observations cover the wavelength region where LyC is expected, we
decided to include this AGN in our final sample. Unfortunately, a
target observed by VLT FORS2 turned out to have a wrong spectroscopic
redshift and we do not considered it in our analysis. For this reason
the final sample contains 16 objects. The AGNs studied in
this paper are summarised in Table \ref{sample}. It is worth noting
here that we systematically avoid observing QSOs classified as Broad
Absorption Line from published spectroscopy, since we expect no LyC
photons escaping from them. As discussed later in Sect. 5, this choice
is not biasing the results reached in the present work.

\begin{figure}
\centering
\includegraphics[width=9cm,angle=0]{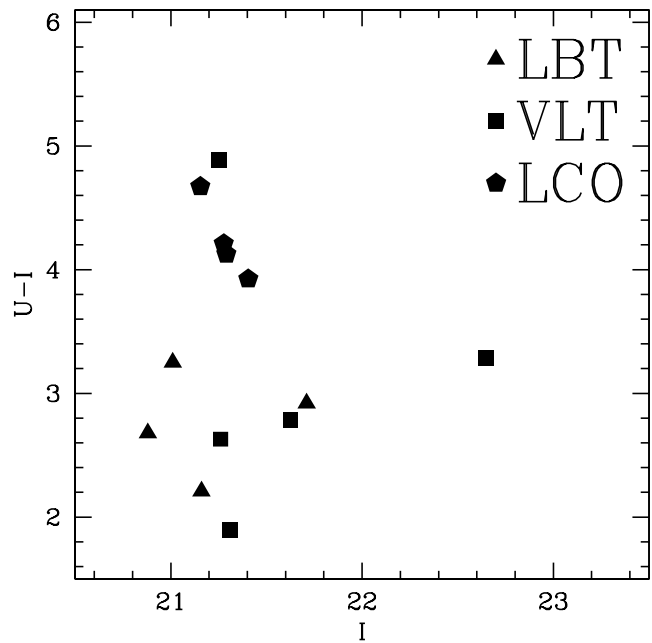}
\caption{The $U-I$ color vs the observed I-band magnitude for the AGNs
studied in this work. Squares show the targets observed with FORS2 at the
VLT telescope, triangles are the AGNs observed with MODS1-2
at the LBT telescope, while pentagons are the AGNs observed with the
LDSS3 instrument at the Magellan-II Clay telescope.
Three objects from Table \ref{sample} have no information on the
U-band magnitude and are not plotted here. We do not plot COSMOS1710
since it is not at redshift $\sim 4$.
}
\label{colormag}
\end{figure}

Fig.\ref{colormag} shows the $U-I$ color versus the observed I-band
magnitude for the AGNs at $3.6<z<4.2$ studied in this paper.
Table \ref{sample} summarises the properties of these faint AGNs.

\begin{table*}
\caption{The AGN sample}
\label{sample}
\centering
\begin{tabular}{l c c c c c c c c}
\hline
\hline
Name & $z_{spec}^{orig}$ & magI & magU & $Log(L_X)$ & Telescope & RA & DEC & $t_{exp}$ \\
     &               & (AB) & (AB) & 2-10 keV &           & J2000 & J2000 & hour \\
\hline
SDSS36 & 4.047 & 21.01 & 24.26 & n.a. & LBT & 01:47:57.46 & +27:33:26.8 & 2.5 \\
SDSS32 & 3.964 & 20.88 & 23.56 & n.a. & LBT & 13:29:26.58 & +28:14:12.9 & 1.75 \\
COSMOS775 & 3.609 & 21.71 & 24.63 & 44.38 & LBT & 09:57:53.49 & +02:47:36.2 & 3.5 \\
SDSS37 & 4.173 & 21.16 & 23.37 & n.a. & LBT & 15:32:36.78 & +27:51:54.8 & 3.5 \\
NDWFSJ05 & 3.900 & 21.95 & -99.0$^{a}$ & n.a. & LBT & 14:36:42.86 & +35:09:23.8 & 5.5 \\
\hline
SDSS04 & 3.772 & 21.31 & 23.21 & n.a. & VLT & 09:32:29.6 & -01:32:32.6 & 1.5 \\
COSMOS1782 & 3.748 & 22.65 & 25.94 & 44.26 & VLT & 10:02:48.9 & +02:22:12.0 & 1.5 \\
SDSS20 & 3.899 & 21.26 & 23.89 & n.a. & VLT & 12:12:48.3 & -01:01:56.3 & 0.83 \\
SDSS27 & 3.604 & 21.63 & 24.41 & n.a. & VLT & 12:49:42.7 & -01:37:22.5 & 0.25 \\
COSMOS955 & 3.715 & 21.25 & 26.14 & 44.71 & VLT & 10:00:50.2 & +02:26:18.5 & 2.17 \\
COSMOS1311 & 3.717 & 21.38 & -99.0$^{a}$ & 44.37 & VLT & 10:01:02.3 & +02:22:34.1 & 2.17 \\
\hline
COSMOS1710 & 3.567$^{b}$ & 22.78 & 23.98 & n.a. & VLT & 10:02:52.1 & +01:55:48.5 & 1.7 \\
\hline
SDSS3777 & 3.723 & 21.28 & 25.49 & n.a. & Magellan & 12:16:44.6 & -01:06:54.4 & 1.5 \\
SDSS3793 & 3.743 & 21.40 & 25.33 & n.a. & Magellan & 12:43:59.8 & -01:59:53.4 & 2.3 \\
SDSS3785 & 3.769 & 21.29 & 25.42 & n.a. & Magellan & 10:30:19.0 & -02:54:56.6 & 1.3 \\
SDSS3832 & 3.663 & 21.15 & 25.83 & n.a. & Magellan & 10:26:31.6 & -00:55:29.7 & 1.3 \\
UDS10275 & 4.096 & 22.27 & -99.0$^{a}$ & n.a. & Magellan & 02:18:05.7 & -05:26:35.8 & 1.7 \\
\hline
\end{tabular}
\\
The spectroscopic redshifts $z_{spec}^{orig}$ have been taken from Dawson et al.
(2013), Glikman et al. (2011), Marchesi et al. (2016), and Civano et al.
(2016). The spectroscopic redshift for UDS10275 has been derived by the Magellan
data described in this paper.
The X-ray luminosity for the AGNs in the COSMOS field has been taken
from Civano et al. (2016).
Notes on individual objects: (a) The U band magnitude for this object is not
available. (b) COSMOS1710 had a spectroscopic redshift
$z_{orig}=3.567$ from the work by Marchesi et al. (2016) and Civano et al.
(2016), but from our deep FORS2 spectrum we derive an updated spectroscopic
redshift of $z=1.547$ (see Fig.\ref{cosmos1710}).
This object is not included in our final sample where
we have measured the LyC escape fraction for AGN at $z\gtrsim 3.6$.
\end{table*}

\subsection{Observations}

We observed 5 objects with 22 hours of exposure in service mode with
the MODS1-2 optical spectrographs at the LBT telescope and 7 targets
with FORS2 (with the blue enhanced CCDs) at the ESO VLT telescope, in
2 nights in visitor mode (Program 098.A-0862, PI A. Grazian). During
several nights at the Magellan-II Clay telescope we observed 4
targets with the LDSS3 spectrograph, and we discovered
one faint AGN at $z\sim 4$ during a pilot project. In total, we
collected a sample of 16 relatively faint AGNs\footnote{We do not
consider here the AGN COSMOS1710 with wrong spectroscopic redshift.}
at $3.6<z<4.2$, which is summarised in Table \ref{sample}. The
adopted exposure time per target varies according to the I band
magnitudes and to the U-I colors of the AGNs.

\subsubsection{LBT MODS}

During the observing period LBT2016 (Program 30; PI A. Grazian), we
carried out deep UV spectroscopy of 5 faint AGNs
(Fig.\ref{colormag}, triangles) down to an absolute magnitude
$M_{1450}=-24.0$ ($L\sim 2 L^*$) at $3.6<z<4.2$ with the LBT double
spectrographs MODS1-2 (Pogge et al. 2012; Rothberg et al. 2016).

MODS is a unique instrument, since it is very efficient in the
$\lambda_{obs}\sim 4000$ {\AA} region, and it allows to observe
in a single exposure the whole optical spectrum, from
$\lambda\sim 3400$ {\AA} to $\sim 10000$ {\AA}, which is essential for
the goal of this paper. We have used MODS1-2, with the
Blue (G400L) and Red (G670L) Low-Resolution gratings fed by the
dichroic, reaching a resolution of $R\sim 1000$ for the adopted slit
of 1.2 arcsec. The dispersion of the spectra is 0.5 {\AA}/px for the
G400L grating in the blue beam, and 0.8 {\AA}/px for the G670L
grating in the red beam, respectively.

The blue side of the MODS1-2 spectra was used to quantify the LyC
escape fraction, sampling the spectral region blue-ward of 912 {\AA}
rest-frame, while the red part ($\lambda\sim 6000-10000$ {\AA}) was
used to fit the continuum at $\lambda_{rest}\sim 1500-2000$ {\AA} of
each AGN with a power law. The simultaneous observations of the blue
and red spectra also allowed to get rid of variability effects which
are affecting the escape fraction studies based on photometry taken in
different epochs (e.g. Micheva et al. 2017). The simultaneous
availability of the blue and the red spectra also allowed us to
increase the survey speed by a factor of 2, with respect to
traditional optical spectrographs on 8m class telescopes (i.e. FORS2
at VLT). Moreover, since the observations were executed in the
``Homogeneous Binocular'' mode (i.e. with MODS1 and MODS2 pointing on
the same position on the sky with the same configuration), by
observing the same target with the two spectrographs the survey speed
gained another factor of 2, for a total net on-target time of 17 hours
for 5 faint AGNs.

The LBT observations were executed in service mode by a dedicated
team under the organization of the LBT INAF Coordination center. The
average seeing during observations was around 1.0 arcsec, with airmass
less than 1.3 and dark moon ($\le 7$ days) in order to go deep in the
UV side of the spectra.

The MODS1-2 spectrographs have been used in long slit spectroscopic mode,
since our targets are sparse in the sky and do not fall on the same
field of view. A dithering of $\pm 1.5$
arcsec along the slit length was carried out in order to improve
the sky subtraction and flat-fielding. The dithered observations were
repeated several times, splitting each observation in sequence of
900 sec exposures.

The relative flux calibration was obtained by observing the
spectro-photometric standard Feige34 for each observing night.
The standard calibration frames (bias, flat, lamps for wavelength
calibration) have been obtained during day-time operation.

\subsubsection{VLT FORS2}

During the observing period ESO P98 (Program 098.A-0862; PI
A. Grazian), we obtained deep FORS2 spectroscopy in visitor
mode for 7 faint AGNs (Fig.\ref{colormag}, squares) down to an
absolute magnitude $M_{1450}=-23.3$ ($L\sim L^*$). Two nights of
VLT (21-22 of February, 2017) were assigned to our program.
We used FORS2 with the blue optimized CCD in visitor mode. Among the
VLT instruments, the blue optimized FORS2 is the {\em only} UV
sensitive instrument that can be used for such scientific application.
We adopted a slit width of 1.0 arcsec with the grism 300V
(without the sorting order filter) which, coupled with the
E2V blue optimized CCD, guarantees
the maximum efficiency in the UV spectral region.

The typical seeing during observations was 0.6-1.4 arcsec, with
partial thin clouds for the majority of the observing run. The
adopted configuration allows us to cover the spectral window from 3400
to 8700 {\AA}, centered at 5900 {\AA}, with a dispersion of 3.4
{\AA}/px and a resolution of $R\sim 400$. Exposure times ranged from 15
minutes to 2.2 hours, depending on the faintness of the targets
and on their $U-I$ color. Each
observation was split in exposures of 1350 seconds, following an ABBA
dithering pattern of $\pm 1.5$ arcsec,
in order to properly subtract the sky background
and to carry out an accurate flat-fielding of the data.

At the beginning of the first night, the spectro-photometric standard
star Hilt600 was observed in order to obtain a relative flux
calibration of the targeted AGNs. All the remaining calibration frames
(bias, flat, lamps for wavelength calibration) were obtained
during day-time operations at the end of each observing night.

Three targets (SDSS04, SDSS20, SDSS27) were observed in long slit
configuration, while the remaining AGNs (COSMOS955, COSMOS1311, COSMOS1710,
COSMOS1782) were observed with multi object spectroscopy (MXU): in the
COSMOS pointings, indeed, there are also many X-ray sources from
Marchesi et al. (2016) and Civano et al. (2016) without spectroscopic
identification. Given the legacy value of these targets, we
carried out MXU observations in order to measure the redshifts of
these fillers.

\subsubsection{Magellan LDSS3}

We used the LDSS3 spectrograph, mounted at the Magellan-II Clay 6.5m
telescope at Las Campanas observatory (LCO), in February and March 2017 to
observe 4 SDSS AGNs with I-band magnitude in the interval $21.0\le
I\le 21.5$. We chose
the grism VPH-Blue, with peak sensitivity around 6000 {\AA},
wavelength coverage between 3800 and 6500 {\AA} and a resolution
$R\sim 1400$ for a slit of 1 arcsec width.
The typical seeing during observations was 0.6-0.9 arcsec, matching the
slit width of 1.0 arcsec. Each
observation was split in exposures of 900 seconds, without any dithering
pattern. The observations were taken
under non-optimal condition w.r.t. the lunar illumination, with a
slightly high background in the UV. Moreover, the sensitivity of
LDSS3 (equipped with red sensitive detectors) drops around 4000 {\AA}
rest frame, where the LyC of $z\sim 4$ sources is expected. For these
reasons, we decided to observe relatively bright candidates in order
to improve the number statistics by enlarging our sample.

In September 2017 a relatively faint AGN (UDS10275, $I\sim 22.3$)
at $z\sim 4.1$ has
been serendipitously observed with the VPH-All grism and the LDSS3
instrument mounted at the Magellan-II Clay 6.5m telescope. The
sensitivity of this grism peaks at $\lambda=7000$ {\AA} and covers the
wavelength range between 4200 and 10000 {\AA}. While the sensitivity
in the blue is slightly reduced w.r.t. the VPH-Blue grism, it allows
to cover a relatively larger spectral window, which is useful for the
characterization of the properties of this target. With an exposure
time of approximately 2 hours we were able to detect flux in the LyC
region, as we will discuss in detail in the following sections.

%__________________________________________________________________

\section{The Method}

\subsection{Data Reduction}

The AGN spectra obtained with MODS1-2 were reduced
by the INAF LBT Spectroscopic Reduction Center based in
Milano\footnote{http://www.iasf-milano.inaf.it/Research/lbt\_rg.html}.

The LBT spectroscopic pipeline was developed inheriting the
functionalities from the VIMOS pipeline (Scodeggio et al. 2005;
Garilli et al. 2012), and was modified for the specific case of
the dual MODS instruments. The pipeline corrects the science frames
with basic pre-reduction (bias, flat field, wavelength calibration in
two dimensions) and subtracts the sky from each frame by fitting the
background with a polynomial function in 2D, which takes
into account the slit distortions. The relative flux calibration was
carried out by computing the spectral response function thanks to
the observations of the spectro-photometric standard star Feige34.
Fully calibrated 2D and 1D extracted spectra
with their associated RMS maps were produced at the end for each
target, stacking all the available LBT observations after the cosmic ray
rejection.

The FORS2 data were reduced with a custom software, developed on the
basis of the MIDAS package (Warmels 1991).
These tools are similar to the one described in Vanzella et
al. (2008). Briefly, the frames have been pre-reduced with the bias
subtraction and the flat fielding. For each slit, the sky background
was estimated with a second order polynomial fitting. The sky fit was
computed independently in each column within two free windows,
above and below the position of the object. In the case of multiple
exposures of the same source, the one
dimensional spectra were co-added weighing according to the
exposure time, the seeing condition and the resulting quality of each
extraction process. Spatial median filtering was applied to each
dithered exposures to eliminate the cosmic rays. Wavelength
calibration was calculated on day-time arc calibration frames, using
four arc lamps (He, HgCd, and two Ar lamps) providing sharp emission
lines over the whole spectral range observed. The object spectra were
then rebinned to a linear wavelength scale.
Relative flux calibration was achieved by observations of the
standard star Hiltner 600.
For each reduced science spectrum we created also an RMS spectrum.

The data acquired with the LDSS3 instrument were reduced using
the same custom package adopted to reduce the FORS2 data, deriving the
relative flux calibrations through the spectro-photometric standard
stars EG274 and LTT4816.

\begin{figure}
\centering
\includegraphics[width=6cm,angle=270]{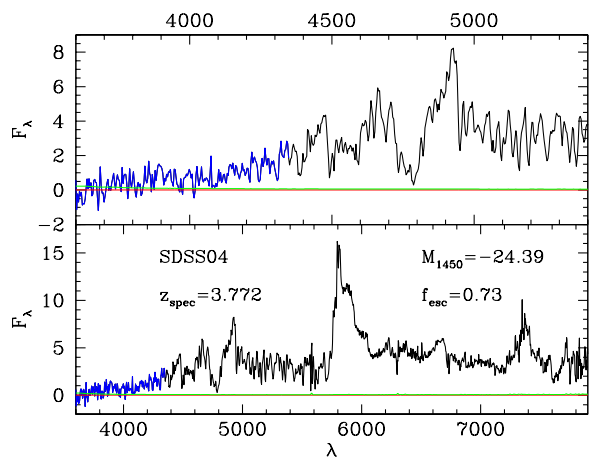}
\caption{{\em Bottom:} The UV/optical spectrum of the AGN SDSS04 observed
by FORS2 at VLT. {\em Top:} A zoom of the blue side of the spectrum for
AGN SDSS04.
The red horizontal lines mark the zero level for the flux
$F_\lambda$, in arbitrary unit. The LyC region (at $\lambda\le 912$
{\AA} rest frame) has been highlighted in blue. The associated RMS is
shown by the green spectrum.
}
\label{sdss04}
\end{figure}

\begin{figure}
\centering
\includegraphics[width=6cm,angle=270]{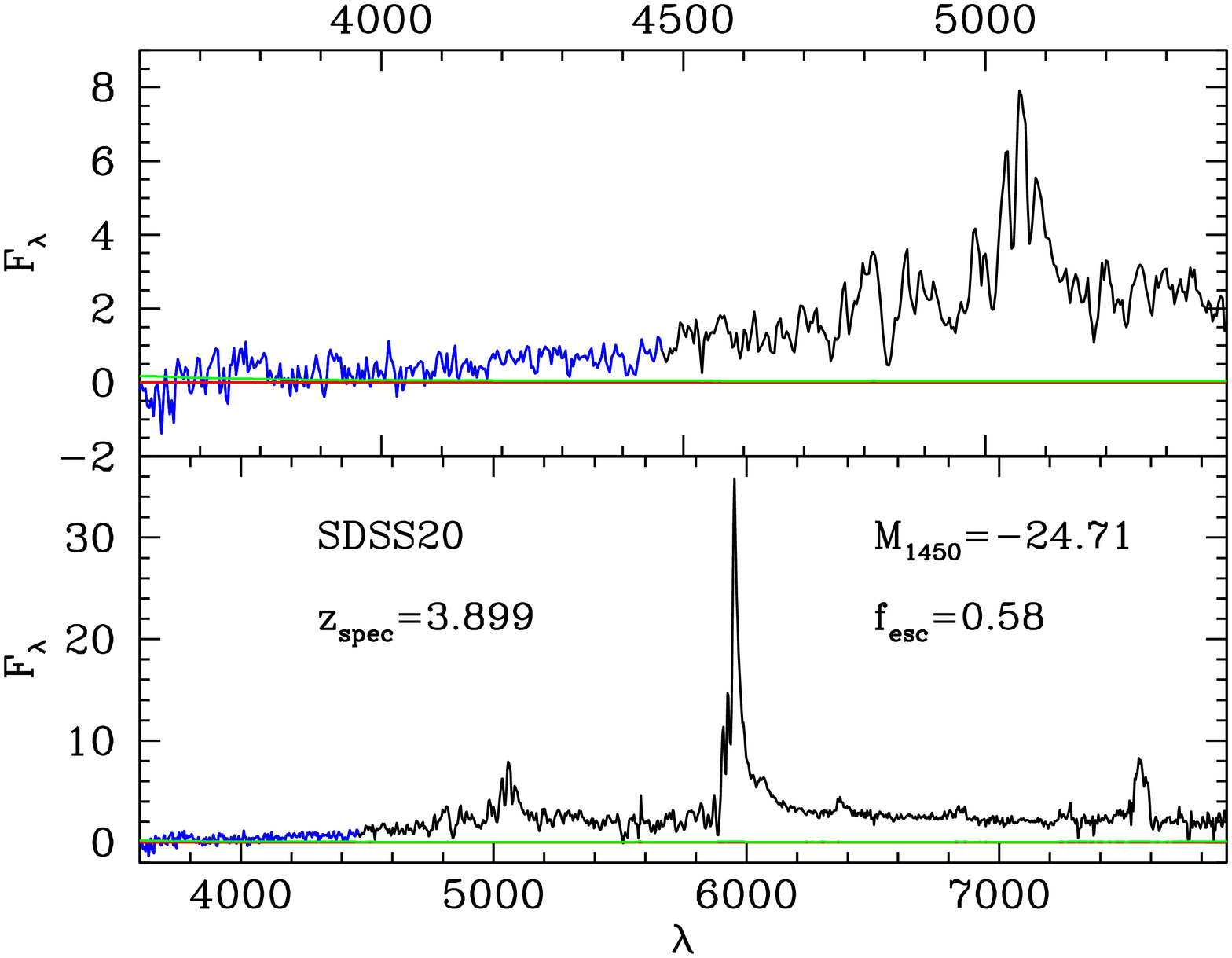}
\caption{{\em Bottom:} The UV/optical spectrum of the AGN SDSS20 observed
by FORS2 at VLT. {\em Top:} A zoom of the blue side of the spectrum for
AGN SDSS20. The red horizontal lines mark the zero level for the flux
$F_\lambda$, in arbitrary unit. The LyC region (at $\lambda\le 912$
{\AA} rest frame) has been highlighted in blue. The associated RMS is
shown by the green spectrum.
}
\label{sdss20}
\end{figure}

\begin{figure}
\centering
\includegraphics[width=6cm,angle=270]{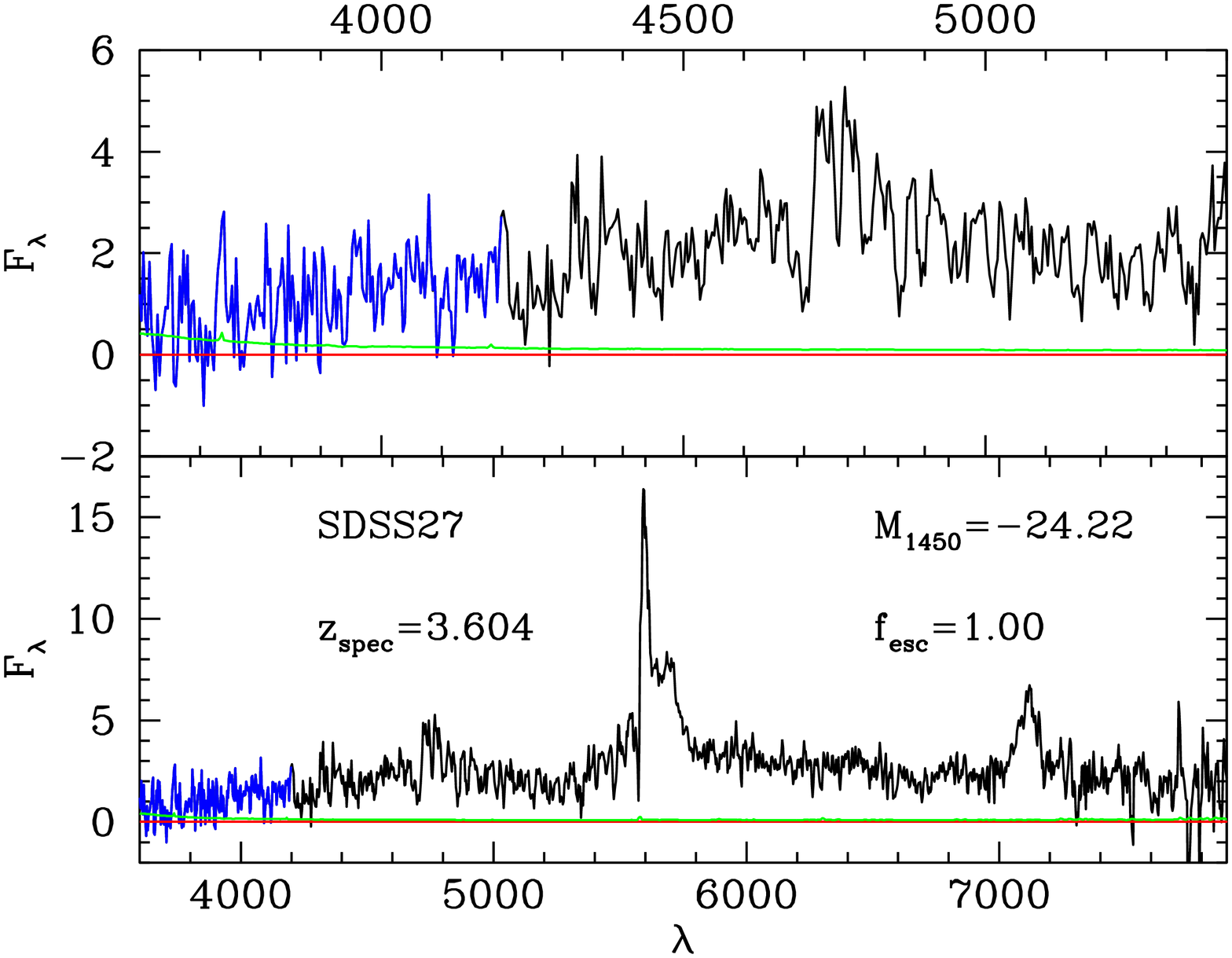}
\caption{{\em Bottom:} The UV/optical spectrum of the AGN SDSS27 observed
by FORS2 at VLT. {\em Top:} A zoom of the blue side of the spectrum for
AGN SDSS27. The red horizontal lines mark the zero level for the flux
$F_\lambda$, in arbitrary unit. The LyC region (at $\lambda\le 912$
{\AA} rest frame) has been highlighted in blue. The associated RMS is
shown by the green spectrum.
}
\label{sdss27}
\end{figure}

\begin{figure}
\centering
\includegraphics[width=6cm,angle=270]{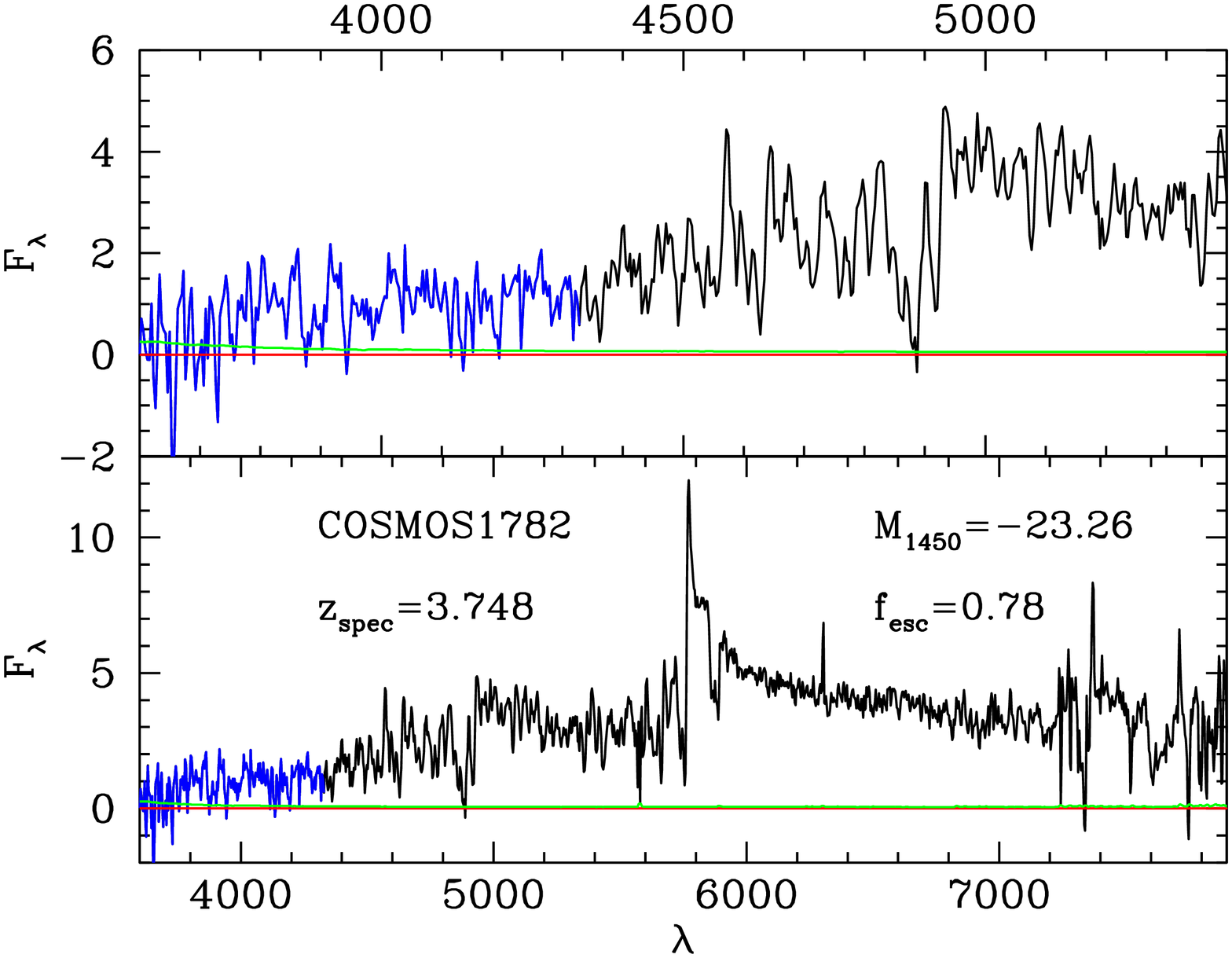}
\caption{{\em Bottom:} The UV/optical spectrum of the AGN COSMOS1782 observed
by FORS2 at VLT. {\em Top:} A zoom of the blue side of the spectrum for
AGN COSMOS1782. The red horizontal lines mark the zero level for the flux
$F_\lambda$, in arbitrary unit. The LyC region (at $\lambda\le 912$
{\AA} rest frame) has been highlighted in blue. The associated RMS is
shown by the green spectrum.
}
\label{cosmos1782}
\end{figure}

\begin{figure}
\centering
\includegraphics[width=6cm,angle=270]{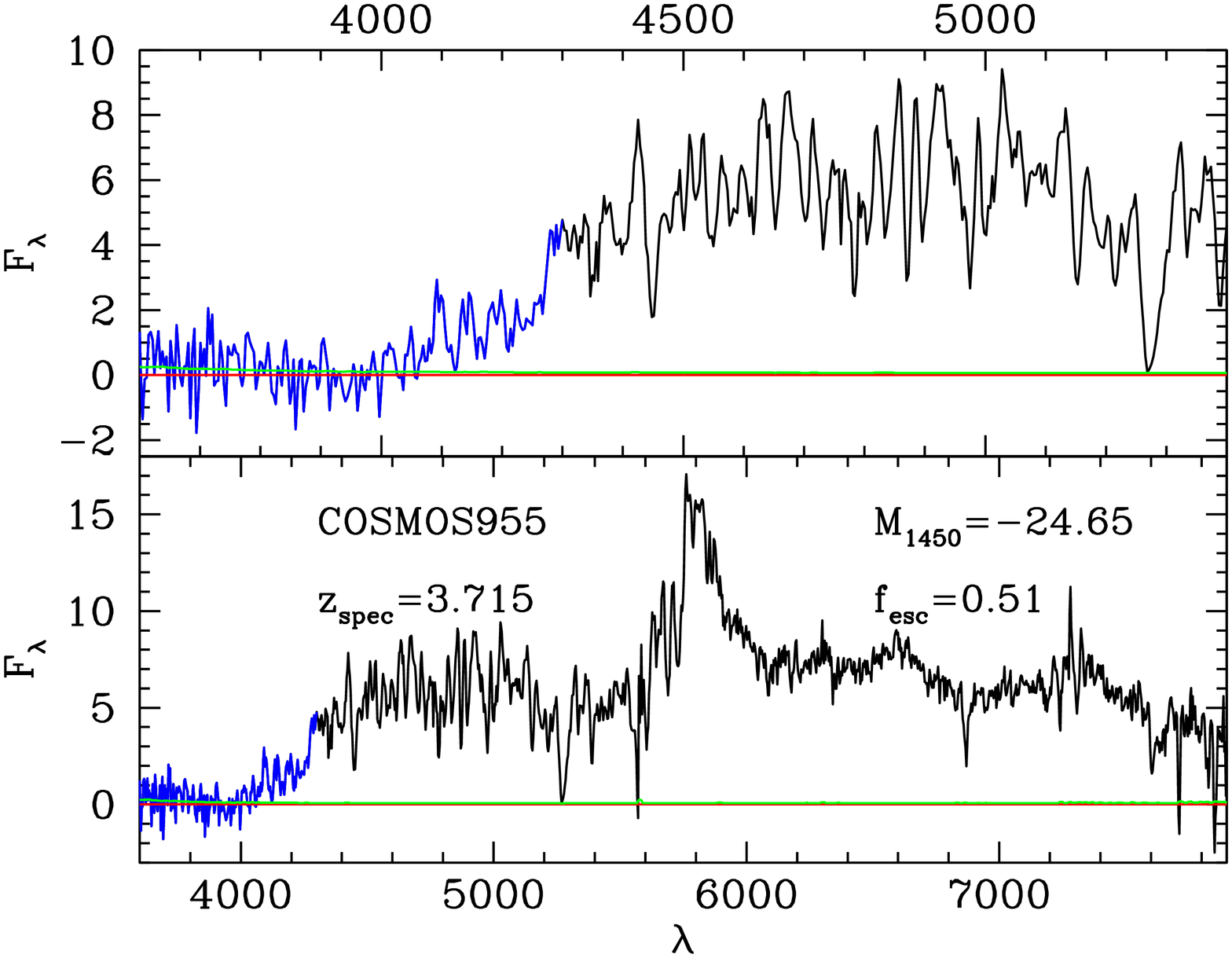}
\caption{{\em Bottom:} The UV/optical spectrum of the AGN COSMOS955 observed
by FORS2 at VLT. {\em Top:} A zoom of the blue side of the spectrum for
AGN COSMOS955. The red horizontal lines mark the zero level for the flux
$F_\lambda$, in arbitrary unit. The LyC region (at $\lambda\le 912$
{\AA} rest frame) has been highlighted in blue. The associated RMS is
shown by the green spectrum.
}
\label{cosmos955}
\end{figure}

\begin{figure}
\centering
\includegraphics[width=6cm,angle=270]{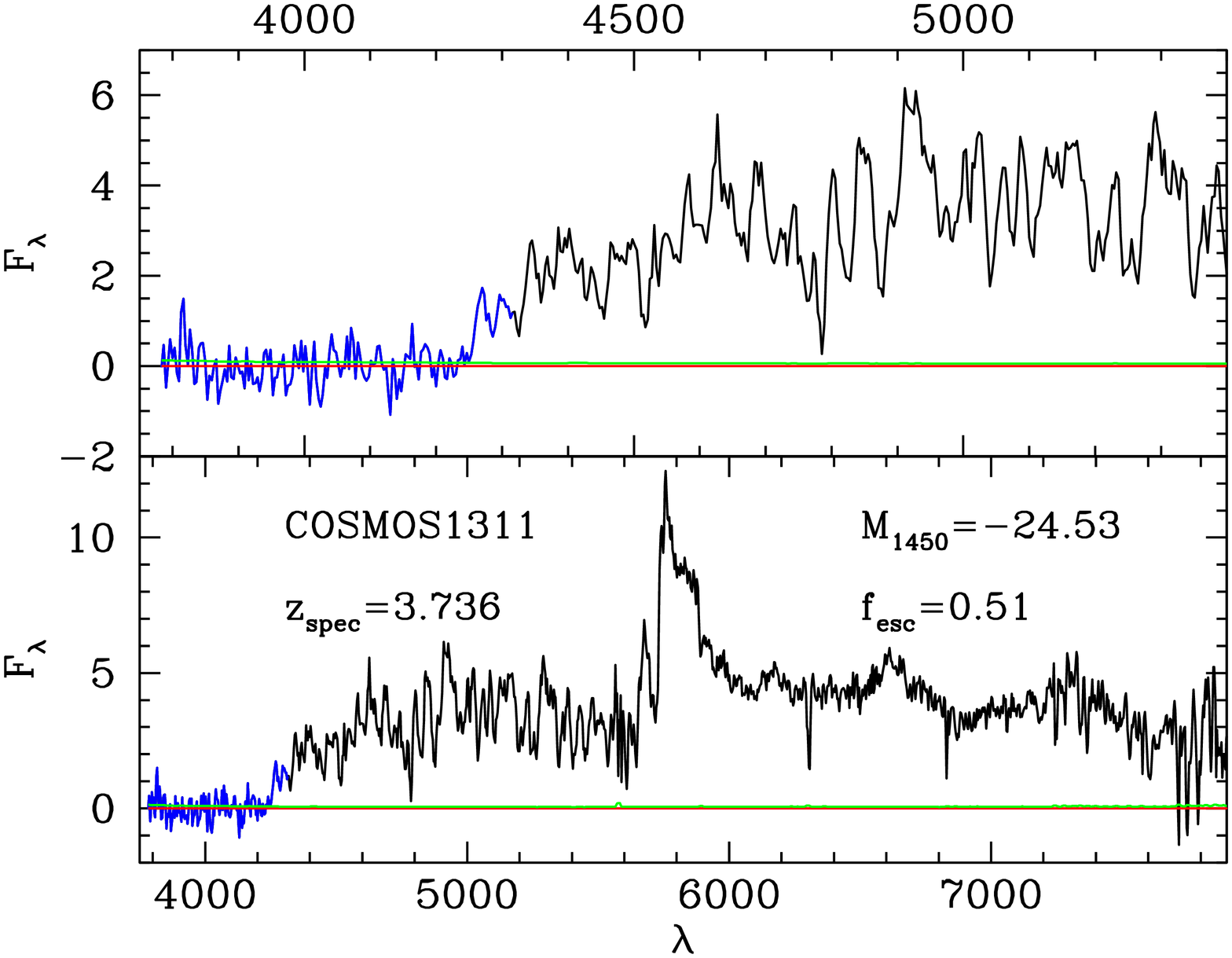}
\caption{{\em Bottom:} The UV/optical spectrum of the AGN COSMOS1311 observed
by FORS2 at VLT. {\em Top:} A zoom of the blue side of the spectrum for
AGN COSMOS1311. The red horizontal lines mark the zero level for the flux
$F_\lambda$, in arbitrary unit. The LyC region (at $\lambda\le 912$
{\AA} rest frame) has been highlighted in blue. The associated RMS is
shown by the green spectrum.
}
\label{cosmos1311}
\end{figure}

\begin{figure}
\centering
\includegraphics[width=6cm,angle=270]{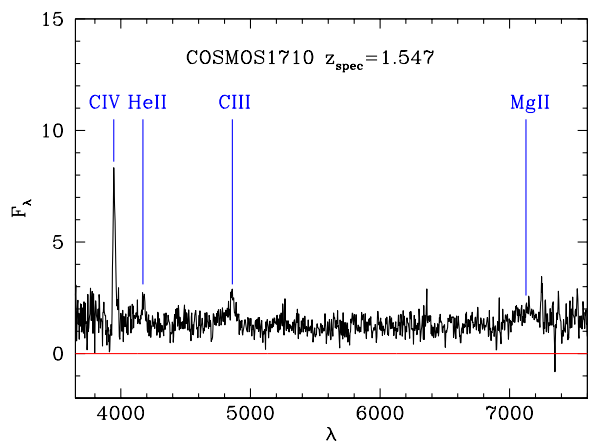}
\caption{The UV spectrum of AGN COSMOS1710 by VLT FORS2.
The spectroscopic redshift for this source in Marchesi et al. (2016) and
Civano et al. (2016) turns out to be
wrong. The correct redshift is $z_{spec}=1.547$,
thanks to many high-ionization lines
detected in the spectrum (CIV, HeII, CIII, MgII, marked by the
blue vertical lines).
}
\label{cosmos1710}
\end{figure}

\begin{figure}
\centering
\includegraphics[width=6cm,angle=270]{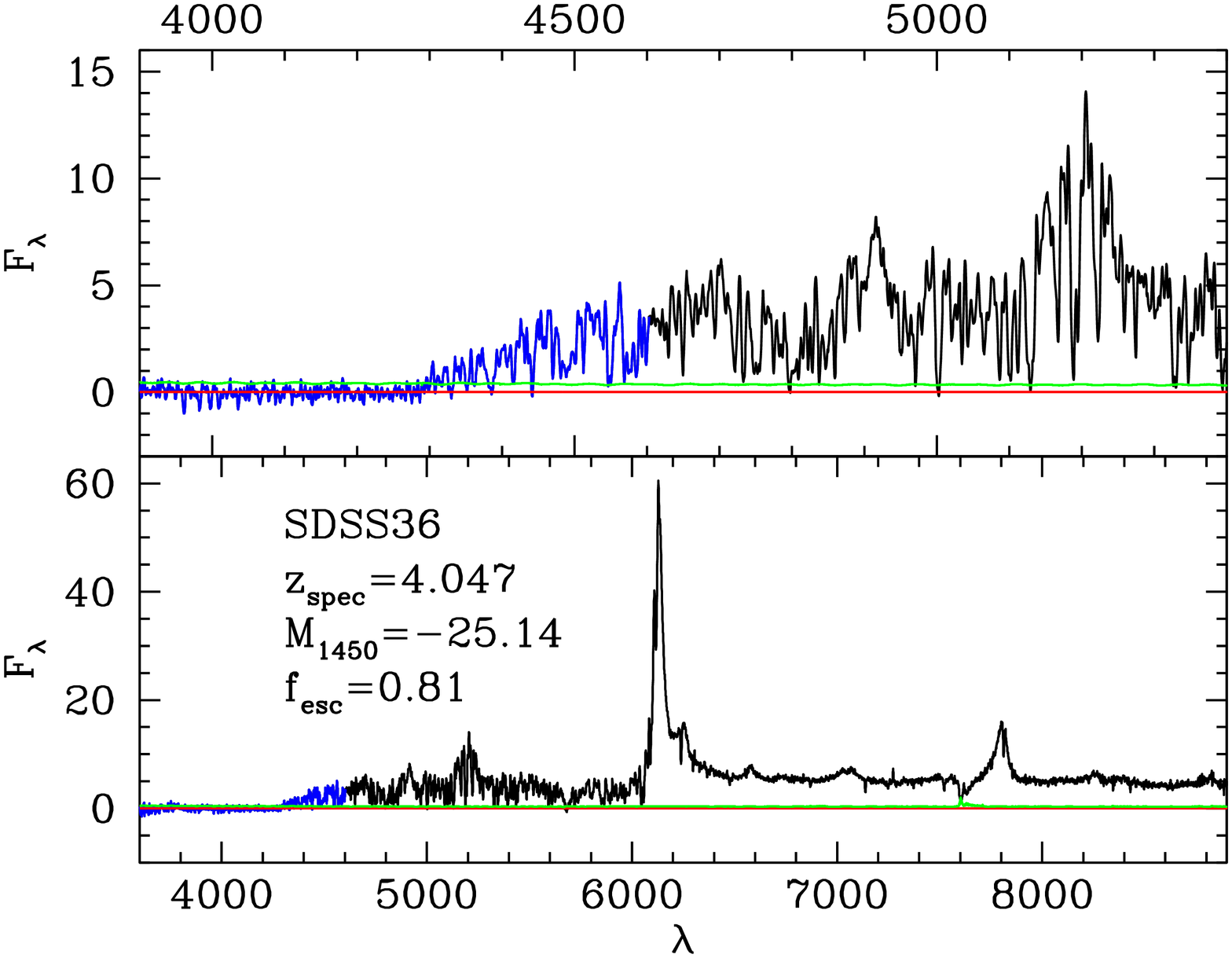}
\caption{{\em Bottom:} The UV/optical spectrum of the AGN SDSS36 observed by
MODS1-2 at LBT. {\em Top:} A zoom of the blue side of the spectrum for
AGN SDSS36.
The red horizontal lines mark the zero level for the flux
$F_\lambda$, in arbitrary unit. The LyC region (at $\lambda\le 912$
{\AA} rest frame) has been highlighted in blue. The associated RMS is
shown by the green spectrum. The scientific data have been smoothed by a
boxcar filter of 5 pixels.
}
\label{sdss36}
\end{figure}

\begin{figure}
\centering
\includegraphics[width=6cm,angle=270]{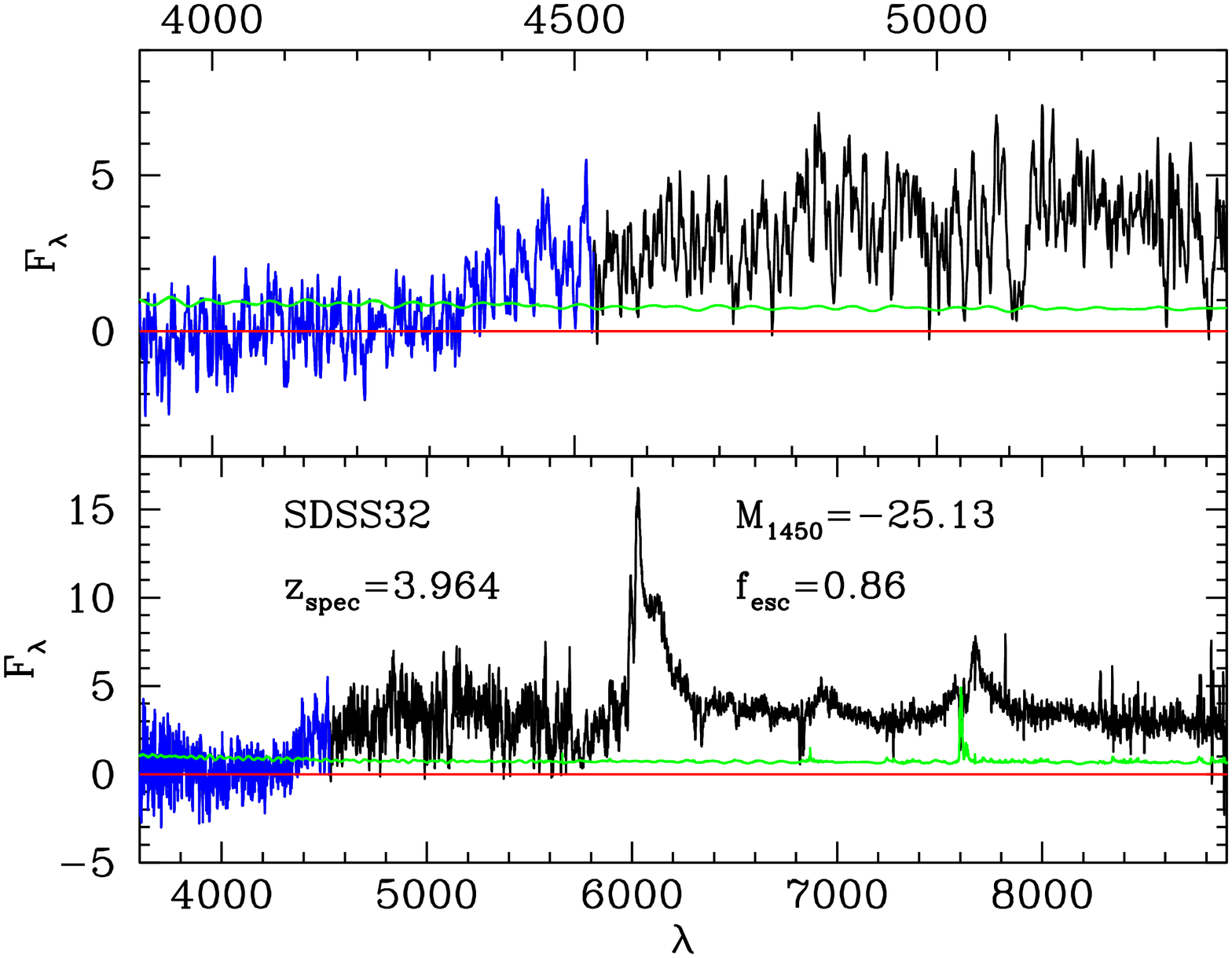}
\caption{The UV/optical spectrum of AGN SDSS32 by LBT MODS1-2.
The LyC region (at $\lambda\le 912$
{\AA} rest frame) has been highlighted in blue. The associated RMS is
shown by the green spectrum. The scientific data have been smoothed by a
boxcar filter of 5 pixels.
}
\label{sdss32}
\end{figure}

\begin{figure}
\centering
\includegraphics[width=6cm,angle=270]{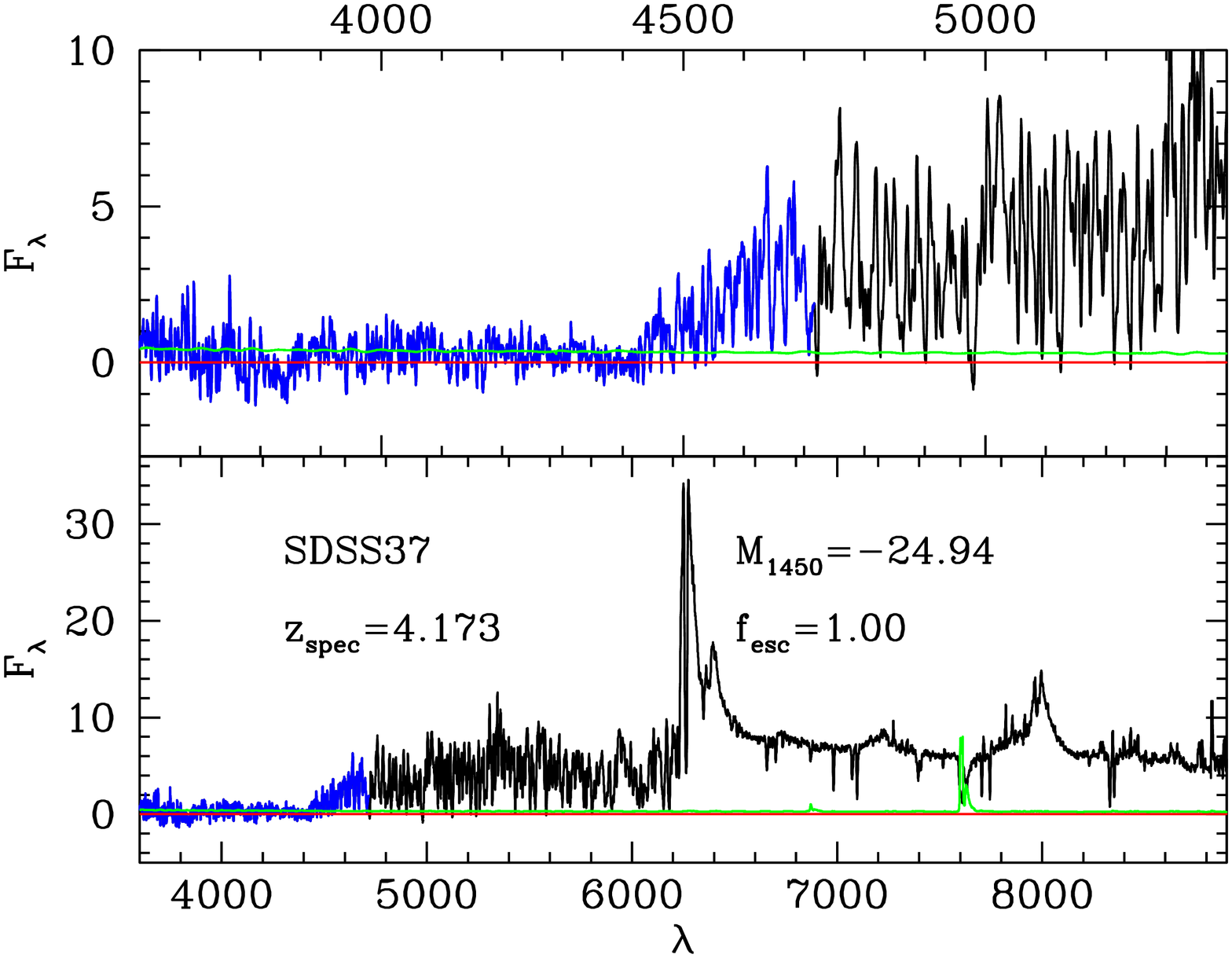}
\caption{The UV/optical spectrum of AGN SDSS37 by LBT MODS1-2.
The LyC region (at $\lambda\le 912$
{\AA} rest frame) has been highlighted in blue. The associated RMS is
shown by the green spectrum. The scientific data have been smoothed by a
boxcar filter of 5 pixels.
}
\label{sdss37}
\end{figure}

\begin{figure}
\centering
\includegraphics[width=6cm,angle=270]{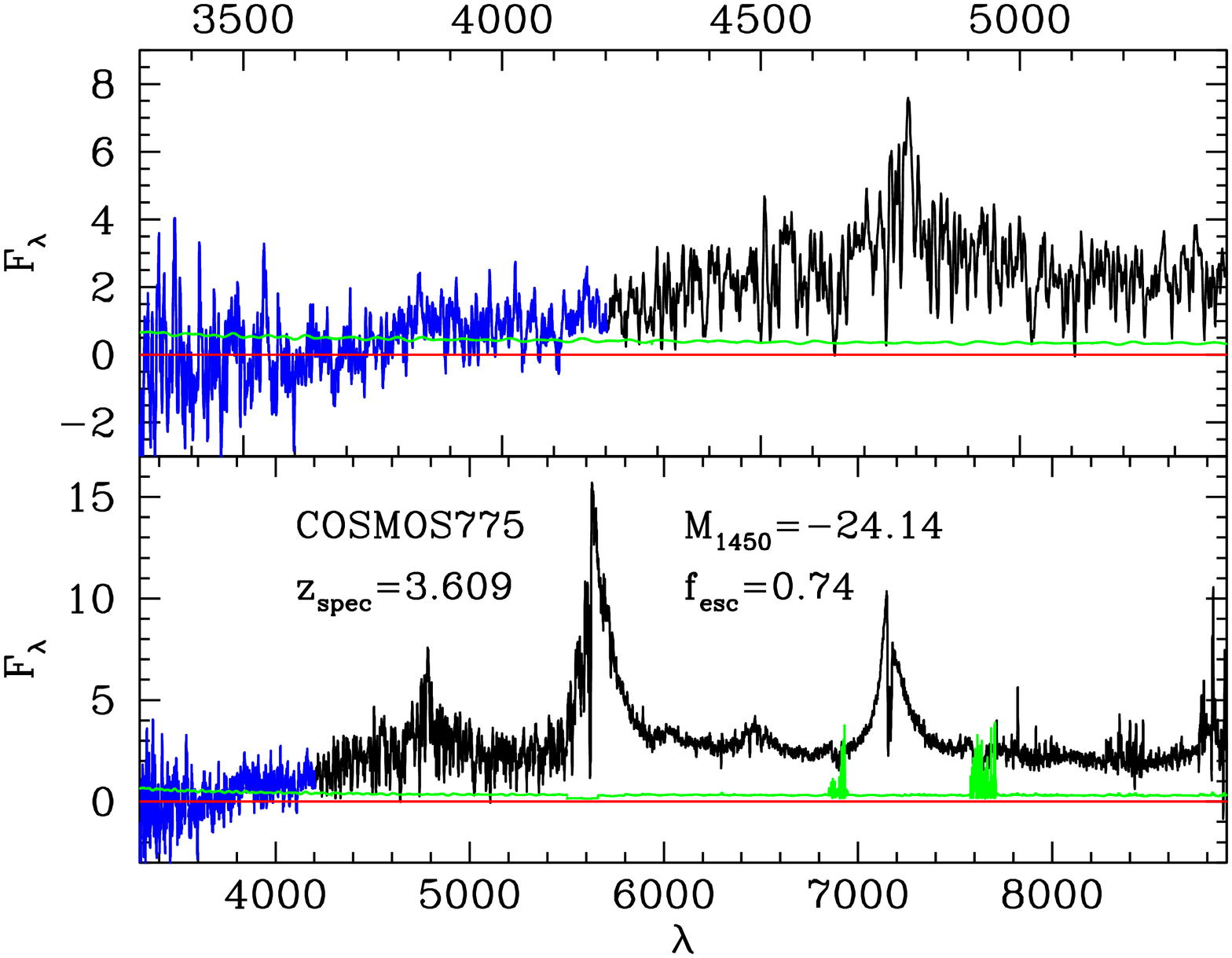}
\caption{The UV/optical spectrum of AGN COSMOS755 by LBT MODS1-2.
The LyC region (at $\lambda\le 912$
{\AA} rest frame) has been highlighted in blue. The associated RMS is
shown by the green spectrum. The scientific data have been smoothed by a
boxcar filter of 5 pixels.
}
\label{cosmos755}
\end{figure}

\begin{figure}
\centering
\includegraphics[width=6cm,angle=270]{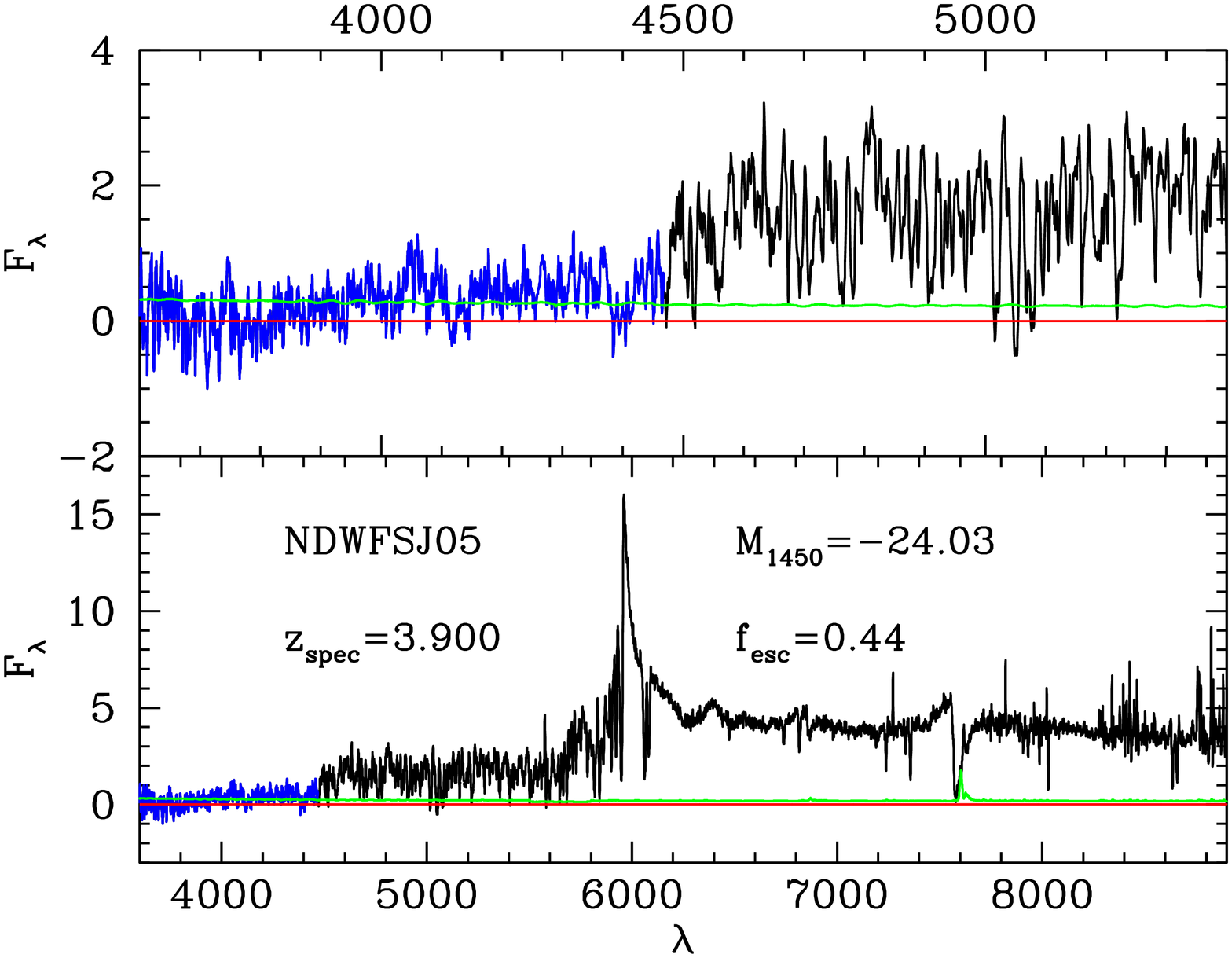}
\caption{The UV/optical spectrum of AGN NDWFSJ05 by LBT MODS1-2.
The LyC region (at $\lambda\le 912$
{\AA} rest frame) has been highlighted in blue. The associated RMS is
shown by the green spectrum. The scientific data have been smoothed by a
boxcar filter of 5 pixels.
}
\label{ndwfsj05}
\end{figure}

\begin{figure}
\centering
\includegraphics[width=6cm,angle=270]{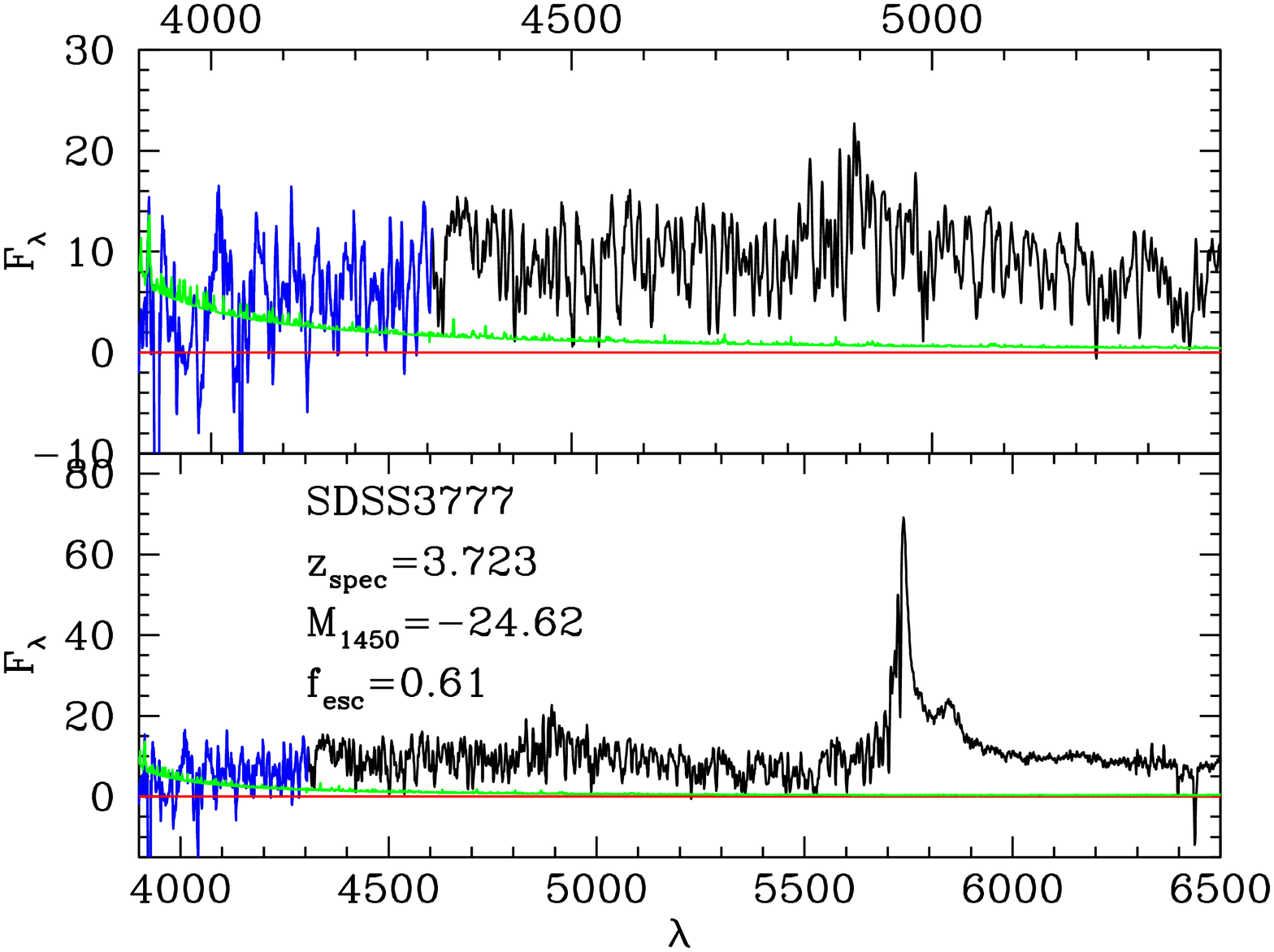}
\caption{{\em Bottom:} The whole spectrum of the AGN SDSS3777 observed by
LDSS3 at Magellan. {\em Top:} A zoom of the blue side of the spectrum for
AGN SDSS3777. The red horizontal lines mark the zero level for the flux
$F_\lambda$, in arbitrary unit. The LyC region (at $\lambda\le 912$
{\AA} rest frame) has been highlighted in blue. The associated RMS is
shown by the green spectrum. The scientific data have been smoothed by a
boxcar filter of 5 pixels.
}
\label{sdss3777}
\end{figure}

\begin{figure}
\centering
\includegraphics[width=6cm,angle=270]{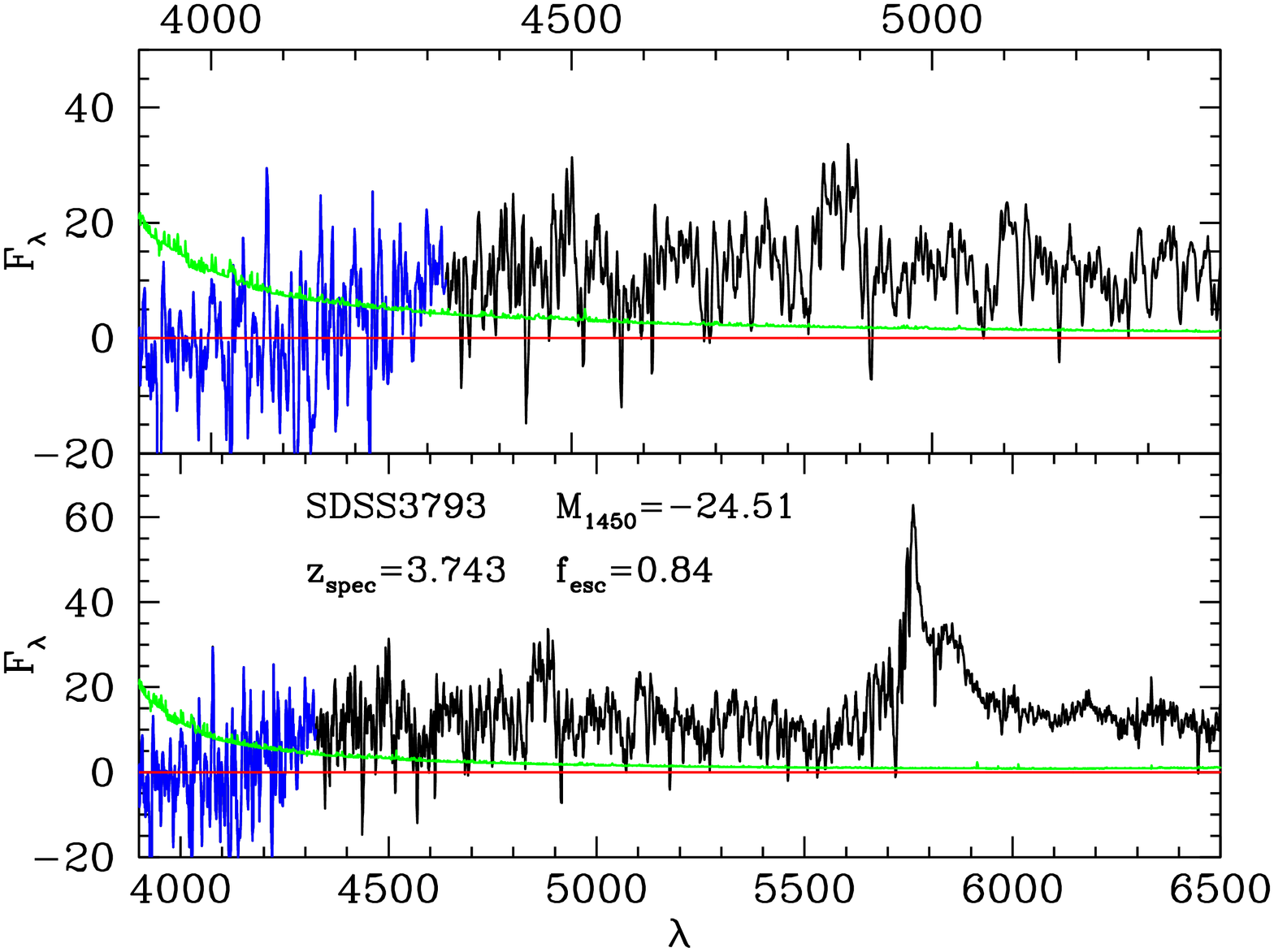}
\caption{{\em Bottom:} The whole spectrum of the AGN SDSS3793 observed by
LDSS3 at Magellan. {\em Top:} A zoom of the blue side of the spectrum for
AGN SDSS3793. The red horizontal lines mark the zero level for the flux
$F_\lambda$, in arbitrary unit. The LyC region (at $\lambda\le 912$
{\AA} rest frame) has been highlighted in blue. The associated RMS is
shown by the green spectrum. The scientific data have been smoothed by a
boxcar filter of 5 pixels.
}
\label{sdss3793}
\end{figure}

\begin{figure}
\centering
\includegraphics[width=6cm,angle=270]{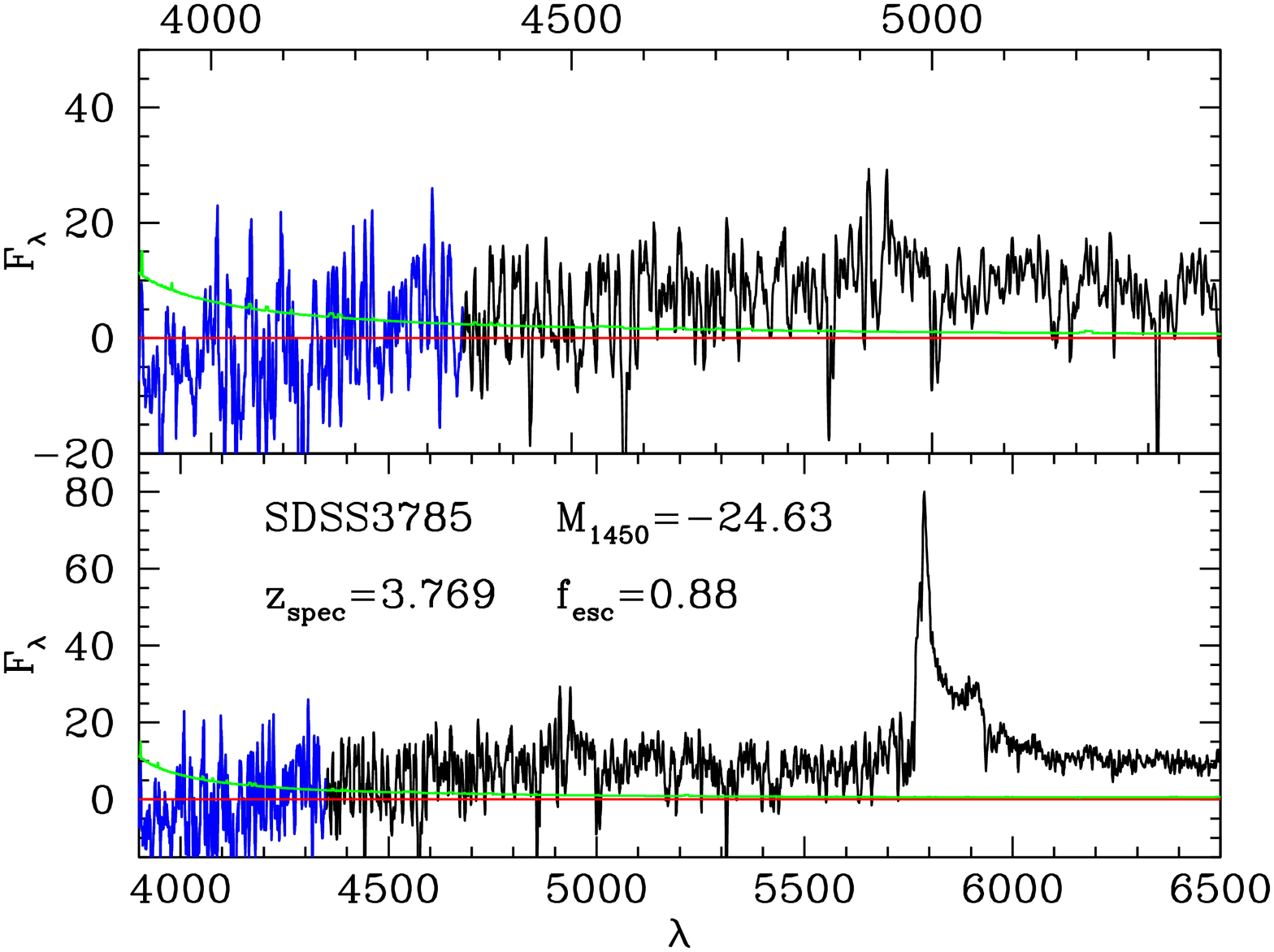}
\caption{{\em Bottom:} The whole spectrum of the AGN SDSS3785 observed by
LDSS3 at Magellan. {\em Top:} A zoom of the blue side of the spectrum for
AGN SDSS3785. The red horizontal lines mark the zero level for the flux
$F_\lambda$, in arbitrary unit. The LyC region (at $\lambda\le 912$
{\AA} rest frame) has been highlighted in blue. The associated RMS is
shown by the green spectrum. The scientific data have been smoothed by a
boxcar filter of 5 pixels.
}
\label{sdss3785}
\end{figure}

\begin{figure}
\centering
\includegraphics[width=6cm,angle=270]{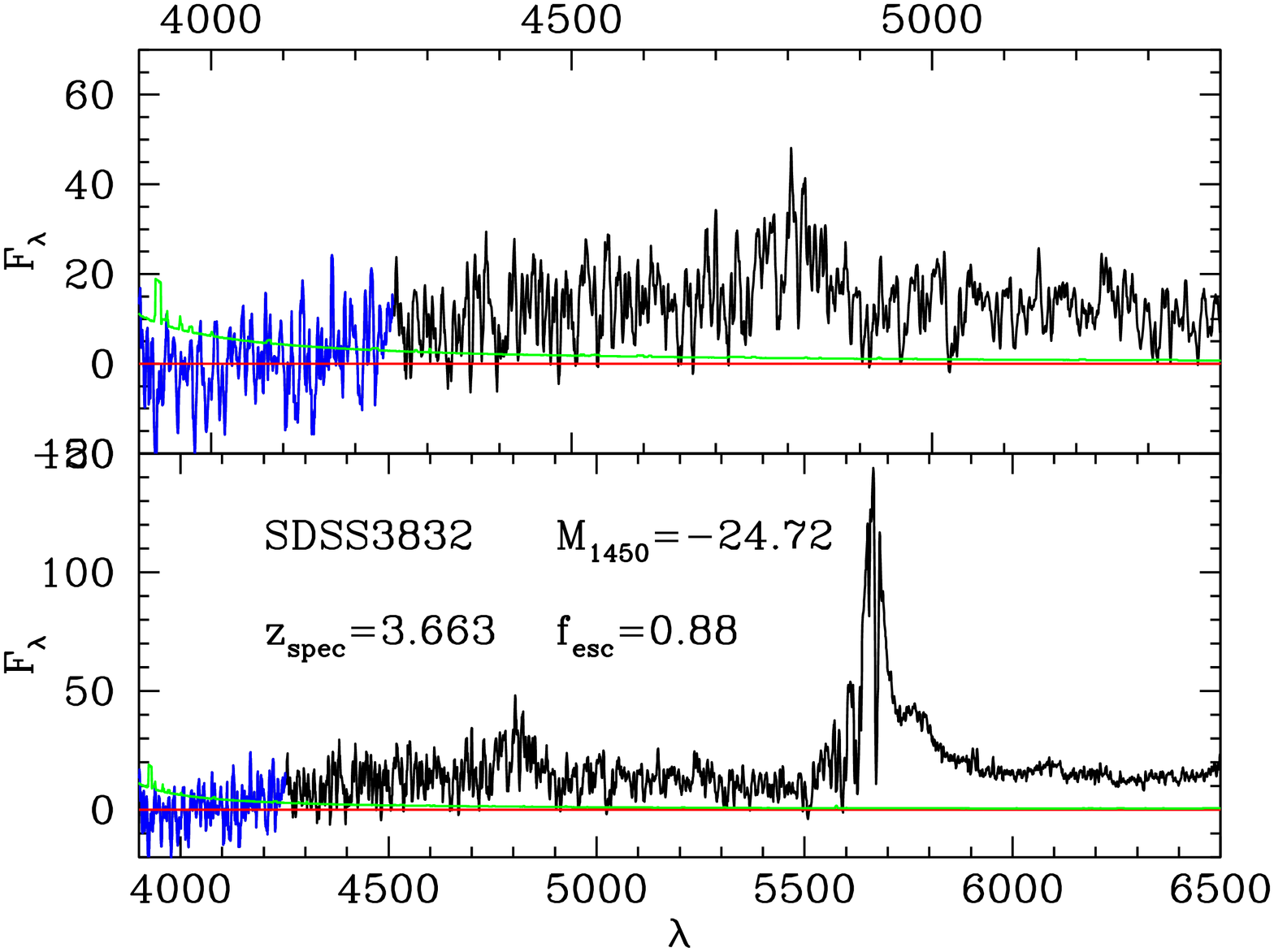}
\caption{{\em Bottom:} The whole spectrum of the AGN SDSS3832 observed by
LDSS3 at Magellan. {\em Top:} A zoom of the blue side of the spectrum for
AGN SDSS3832. The red horizontal lines mark the zero level for the flux
$F_\lambda$, in arbitrary unit. The LyC region (at $\lambda\le 912$
{\AA} rest frame) has been highlighted in blue. The associated RMS is
shown by the green spectrum. The scientific data have been smoothed by a
boxcar filter of 5 pixels.
}
\label{sdss3832}
\end{figure}

\begin{figure}
\centering
\includegraphics[width=6cm,angle=270]{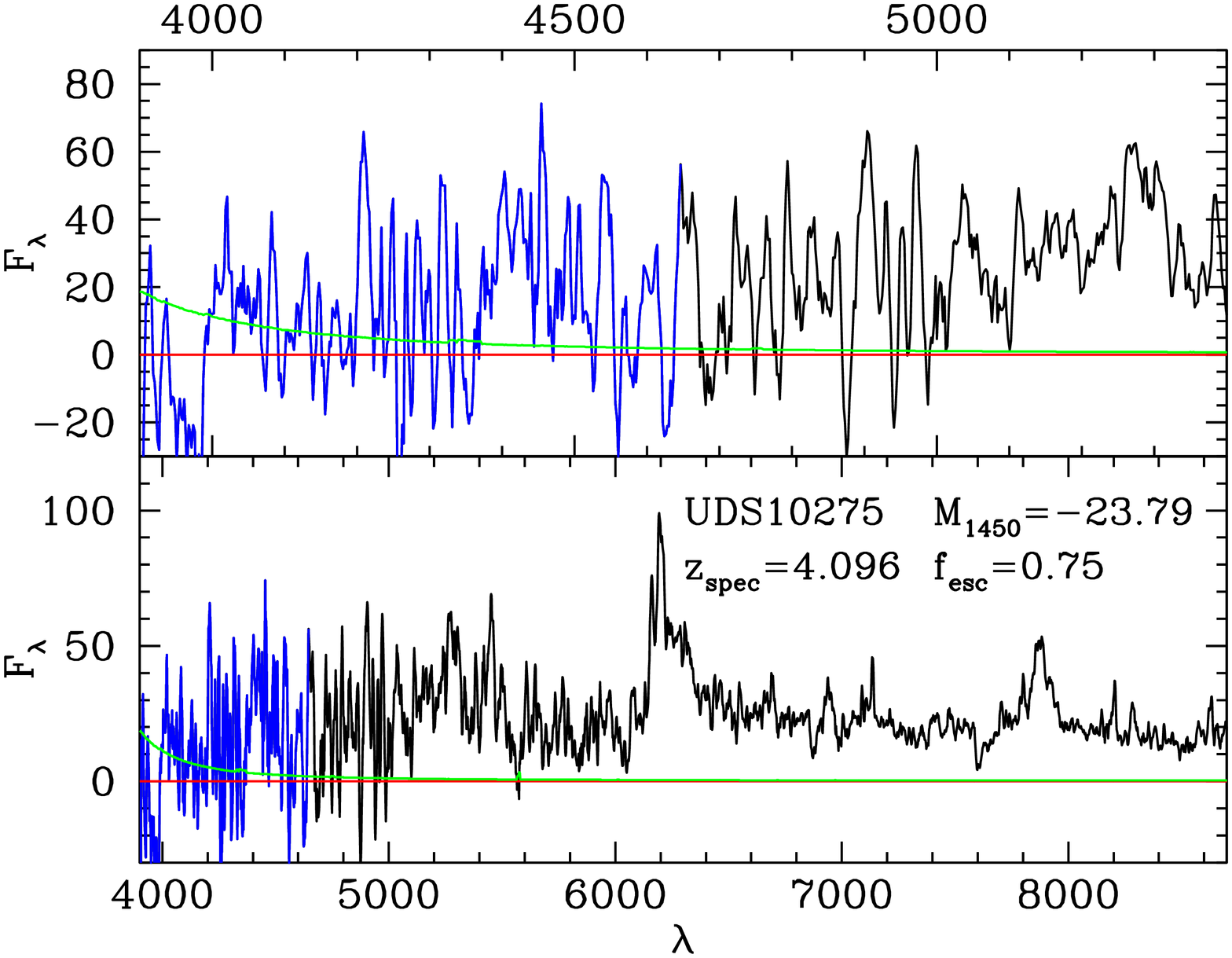}
\caption{{\em Bottom:} The whole spectrum of the AGN UDS10275 observed by
LDSS3 at Magellan. {\em Top:} A zoom of the blue side of the spectrum for
AGN UDS10275. The red horizontal lines mark the zero level for the flux
$F_\lambda$, in arbitrary unit. The LyC region (at $\lambda\le 912$
{\AA} rest frame) has been highlighted in blue. The associated RMS is
shown by the green spectrum. The scientific data have been smoothed by a
boxcar filter of 5 pixels.
}
\label{uds10275}
\end{figure}

Fig.\ref{sdss04}-\ref{cosmos1710}, Fig.\ref{sdss36}-\ref{ndwfsj05},
and Fig.\ref{sdss3777}-\ref{uds10275} show the AGN spectra obtained
with FORS2 at VLT, MODS1-2 at LBT, and LDSS3 at Magellan telescope,
respectively, as described in Table \ref{sample}. In each spectrum,
the blue region shows the spectral window covering LyC emission,
i.e. at $\lambda\le 912$ {\AA} rest frame. The Magellan and LBT
spectra have been smoothed by a boxcar filter of 5 pixels for
aesthetic reasons only, in order to match the spectral resolution with
the effective resolution of the pictures.

\subsection{Estimating the Lyman Continuum escape fraction of the faint AGNs}

For each object in Table \ref{sample} the following measurements have
been carried out. First, we refine the input spectroscopic redshift by
measuring the position of the OI 1305 emission line, which gives the
systemic redshift. When this line is weak, we keep the original
redshift $z_{spec}^{orig}$ provided in Table \ref{sample}, otherwise the
updated redshift $z_{spec}^{new}$ can be found in Table \ref{tabresu}.
The exact determination of the systemic redshift for our
AGNs is important since it allows us to measure precisely the position
of the 912 {\AA} break, and thus an accurate estimate of the LyC
escape fraction for our objects. Only for the AGNs COSMOS1311 and SDSS04 the
spectroscopic redshifts have been revised by small amounts,
$\Delta z\sim 0.019$ and $-0.004$ respectively.
The AGN COSMOS1710 instead had a wrong
spectroscopic redshift in Marchesi et al. (2016) and Civano et
al. (2016) and was discarded in the following analysis.

\begin{figure}
\centering
\includegraphics[width=6cm,angle=270]{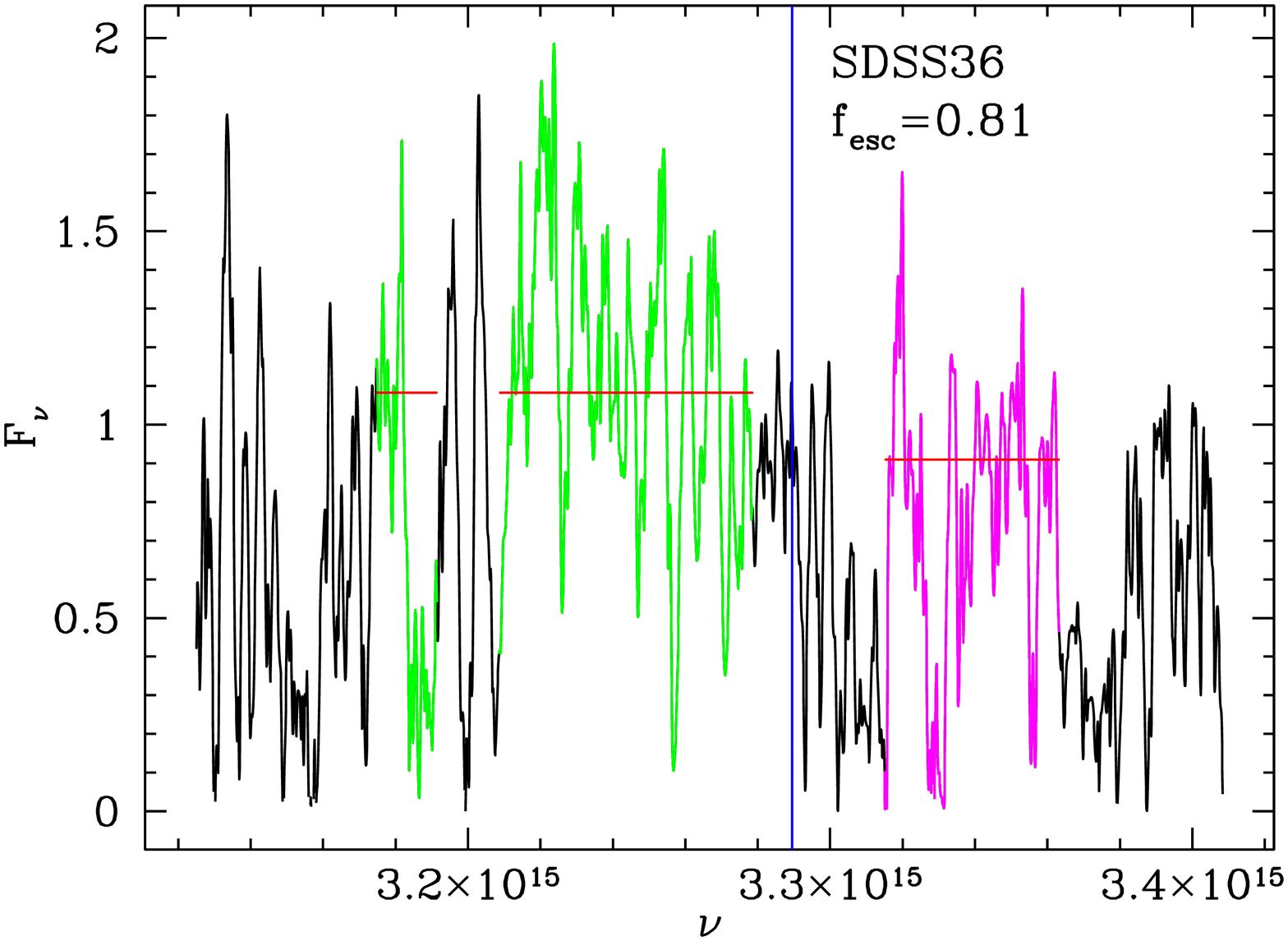}
\caption{
The estimate of the LyC escape fraction for the AGN SDSS36 observed
with MODS1-2 at the LBT telescope. The spectrum is in $F_\nu$ and it
is arbitrarily normalized and blueshifted to $z=0$ (rest frame frequencies).
The green portion of the spectrum shows the
spectral region between 915 and 945 {\AA} rest frame (excluding the
wavelength range 935-940 {\AA} due to the presence of the
Lyman-$\epsilon$ emission line), while the magenta portion indicates
the ionizing photons emitted between 892 and 905 {\AA} rest frame.
The blue vertical line indicates the location of the 912 {\AA} rest
frame break. The red horizontal lines mark the mean values above and
below the Lyman limit, after the iterative 2-$\sigma$ clipping. The
resulting escape fraction is the ratio between these two mean fluxes,
and turns out to be $81\%$ for SDSS36.
}
\label{fesc}
\end{figure}

After the refinement of the spectroscopic redshifts, we compute for each
AGN in our sample the LyC escape fraction. We decided to adopt the
technique outlined in Sargent et al. (1989) in order to measure
$f_{esc}=exp(-\tau_{LL})$ from the spectra, where $\tau_{LL}$ is the opacity of
the associated Lyman limit (LL). Precisely, we estimate the
mean flux above and below the Lyman limit (912 {\AA} rest frame) and
measure the escape fraction as $f_{esc}=f_\nu(900)/f_\nu(930)$, where
$f_\nu(900)$ is the mean flux of the AGN in the Lyman continuum region,
namely between 892 and 905 {\AA} rest frame, while $f_\nu(930)$ is the
average flux in the non ionizing region redward of the LL,
between 915 and 945 {\AA} rest frame, avoiding the region between 935 and
940 {\AA} due to the presence of the Lyman-$\epsilon$ emission line.
These average fluxes have been computed through an iterative clipping
of spectral regions deviating more than 2 $\sigma$ from the mean flux values,
as shown in Fig.\ref{fesc} for AGN SDSS36.
This method allows us to avoid spectral regions affected
by some intervening strong IGM absorption systems or contaminated
by emission lines. Similarly to Prochaska et al. (2009),
we do not use the wavelength range close to the Lyman limit (905-912 {\AA}
rest frame) since it can be affected by the AGN proximity effect.
In principle, the AGN proximity zone is a signature of ionizing photons
escaping into the IGM, and should be considered in these calculation.
However, we do not want to work too
close to 912 {\AA} rest frame in order to avoid biases due to possible
wrong estimates of the spectroscopic redshift. For this reason the
LyC $f_{esc}$ provided in Table \ref{tabresu}
should be considered as a robust lower limit for the real value.

This technique is different compared to the method adopted by Cristiani et
al. (2016). They fit the SDSS QSO spectra with a power law in the
wavelength range between 1284 and 2020 {\AA} rest frame, avoiding the
region affected by strong emission lines, and then extrapolate the
fitted spectrum blueward of the Lyman-$\alpha$ line. After correcting
for the spectral slope, they apply a mean correction for IGM
absorption adopting the recipes of Inoue et al. (2014). Finally they
computed the escape fraction as the mean flux between 865 and 885
{\AA} rest frame, after normalizing the spectra to 1.0 redward of the
Lyman-$\alpha$. Since they apply an average correction for the IGM
extinction, the LyC escape fraction of their bright QSOs goes from 0.0
to 2.5 (250\%, see their Fig. 7). The values above 100\% are simply
due to the fact that in some QSOs the actual IGM absorption is lower than the
mean value provided by Inoue et al. (2014). Since we do not want to
rely on assumptions regarding the IGM properties surrounding our
AGNs, we decided to adopt the technique of Sargent et al. (1989)
outlined above, which gives a robust lower limit for the LyC escape fraction.

Finally we computed the absolute magnitude of our AGNs at 1450 {\AA}
rest frame starting from the observed I band magnitude in Table \ref{sample},
$M_{1450}=I-5\log_{10}(D_L)+5+2.5\log_{10}(1+z_{spec})$,
where $D_L$ is the
luminosity distance of the object and $z_{spec}$ is the refined
spectroscopic redshift. At $z\sim 4$ the I band is sampling directly
the 1450 {\AA} rest frame, thus minimizing the K-correction effects.

Table \ref{tabresu} summarises the measured properties (refined
spectroscopic redshift, LyC escape fraction, absolute magnitude
$M_{1450}$) of the faint AGNs in our sample.

\begin{table}
\caption{The measured properties of faint AGNs in our sample}
\label{tabresu}
\centering
\begin{tabular}{l c c c c}
\hline
\hline
Name & $z_{spec}^{new}$ & $f_{esc}(LyC)$ & S/N & $M_{1450}$ \\
\hline
SDSS36 & 4.047 & 0.81 & 87 & -25.14 \\
SDSS32 & 3.964 & 0.86 & 33 & -25.13 \\
COSMOS775 & 3.609 & 0.74 & 31 & -24.14 \\
SDSS37 & 4.173 & 1.00 & 121 & -24.94 \\
NDWFSJ05 & 3.900 & 0.44 & 12 & -24.03 \\
\hline
SDSS04 & 3.768 & 0.73 & 96 & -24.39 \\
COSMOS1782 & 3.748 & 0.78 & 72 & -23.26 \\
SDSS20 & 3.899 & 0.53 & 58 & -24.71 \\
SDSS27 & 3.604 & 1.00 & 42 & -24.22 \\
COSMOS955 & 3.715 & 0.51 & 84 & -24.65 \\
COSMOS1311 & 3.736 & 0.51 & 29 & -24.53 \\
\hline
SDSS3777 & 3.723 & 0.61 & 26 & -24.62 \\
SDSS3793 & 3.743 & 0.84 & 12 & -24.51 \\
SDSS3785 & 3.769 & 0.88 & 20 & -24.63 \\
SDSS3832 & 3.663 & 0.88 & 11 & -24.72 \\
UDS10275 & 4.096 & 0.75 & 27 & -23.80 \\
\hline
\hline
MEAN & 3.82 & 0.74 & & -24.46 \\
\hline
\hline
\end{tabular}
\\
The LyC escape fraction $f_{esc}(LyC)$ and the absolute
magnitude $M_{1450}$ have been
derived adopting the refined spectroscopic redshift $z_{spec}^{new}$.
The S/N ratio refers to the total flux in the LyC region, integrated between
892 and 905 {\AA} rest frame. The errors on $f_{esc}(LyC)$ vary from $\sim$2
to 15\%, according to the measured S/N ratio.
\end{table}

%-----------------------------------------------------------------

\section{Results on the LyC escape fraction of high-z AGNs}

The results summarised in Table \ref{tabresu} indicate that we detect
a LyC escape fraction between 44\% and 100\% for all the 16 observed AGNs
with absolute magnitude in the range $-25.14\le M_{1450}\le
-23.26$. From a quantitative analysis of the spectra shown in
Fig.\ref{sdss04} and \ref{sdss36}, as well as those in the Appendix, we
confirm that the detection of HI ionizing flux is significant for all
the observed AGNs, with a S/N ratio between 11 and 121 for our
targets. The uncertainties on the measured LyC escape
fraction in Table \ref{tabresu} are of few percents ($\sim 2-15\%$).
The very good quality of these data has been allowed by the
usage of efficient spectrographs in the UV wavelengths and by the long
exposure time dedicated to this program (see Table \ref{sample}).
For the AGNs observed with the Magellan-II telescope the S/N is slightly
lower, S/N$\sim 11-27$, due to the non optimal conditions during the
observations (high moon illumination).

With these spectra we confirm the detection of ionizing radiation for
all the 16 observed AGN, with a mean LyC escape fraction of $\sim
74\%$ and a dispersion of $\pm 18\%$ at 1$\sigma$ level. The latter
should not be considered as the uncertainty on the $f_{esc}$
measurements, but the typical scatter of the observed AGN sample.

\subsection{Dependence of LyC escape fraction on AGN luminosity and U-I color}

\begin{figure}
\centering
\includegraphics[width=9cm,angle=0]{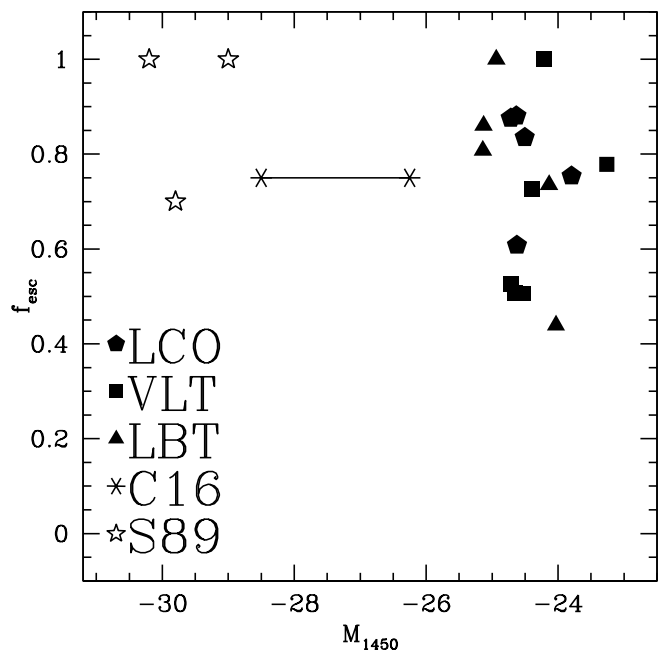}
\caption{
The dependence of the LyC escape fraction on the absolute magnitude
$M_{1450}$ for QSOs and AGNs at $3.6\le z\le 4.2$. Squares show the
targets observed with the VLT telescope, triangles are the AGNs
observed with MODS1-2 at the LBT telescope, while pentagons are the
AGNs observed with the LDSS3 instrument at the Magellan-II Clay
telescope. The uncertainties on the measured LyC escape fraction at
$M_{1450}\gtrsim -25$ are of few percents ($\sim 2-15\%$). The two
asterisks connected by an horizontal line show the range of $M_{1450}$
for the QSOs studied by Cristiani et al. (2016, C16), while the three
stars are three very bright QSOs ($M_{1450}\sim -30$) studied by
Sargent et al. (1989, S89). No obvious trend of $f_{esc}$ with the
absolute magnitude $M_{1450}$ is detected.
}
\label{fescmabs}
\end{figure}

Fig.\ref{fescmabs} shows the dependence of the escape fraction
on the absolute magnitude $M_{1450}$ of the faint AGNs in our sample
(filled triangles, squares, and pentagons).
No particular trend with the absolute
magnitude is observed in our data. In
order to extend the baseline for the AGN luminosities, we adopt as
reference the mean value of the escape fraction derived by Cristiani
et al. (2016) for a sample of 1669 bright QSOs at $z\sim 4$ from the
SDSS survey. They obtain a mean escape fraction of 75\% for
$M_{1450}\lesssim -26$, which is a luminosity $\sim$10 times larger than our
faint limit $M_{1450}\le -23.26$. As evident from Fig.\ref{fescmabs},
no trend of the escape fraction of ionizing photons with the
luminosity of the AGN/QSOs is detected. It is interesting to note that
the quoted value for Cristiani et al. (2016) of $M_{1450}\lesssim -26$
represents the faint limit of the 1669 SDSS
QSOs analysed, and their sample extends towards brighter limits
$M_{1450}\sim -29$.
Moreover, in Sargent et al. (1989) there are two
QSOs (Q0000-263 and Q0055-264) with redshift $z>3.6$ and absolute
magnitude of -29.0 and -30.2, with escape fraction close to 100\%,
and one QSO (Q2000-330) with $M_{1450}=-29.8$ and $f_{esc}\sim 70\%$.
Finally, it is worth noting that the value of escape fraction provided
in Table \ref{tabresu} are bona fide
lower limits to the ionizing radiation, which can be even higher than 80\%
and possibly close to 100\%, if we take into account all the possible
corrections (i.e., proximity effect, absorbers close to the LLS, intrinsic
spectral slopes), as we will discuss in Section 5.

The outcome of Fig.\ref{fescmabs} is that the LyC escape
fraction of QSOs and their fainter version (AGNs) does not vary in a
wide luminosity range, between $M_{1450}\sim -30$ down to
$M_{1450}\sim -23$, which is a factor of $10^{3}$ in luminosity. More
interestingly, we are reaching luminosities $\lesssim L^*$ at $z\sim 4$,
which should provide the bulk of the emissivity at 1450 {\AA}
rest frame (Giallongo et al. 2012, 2015).

\begin{figure}
\centering
\includegraphics[width=9cm,angle=0]{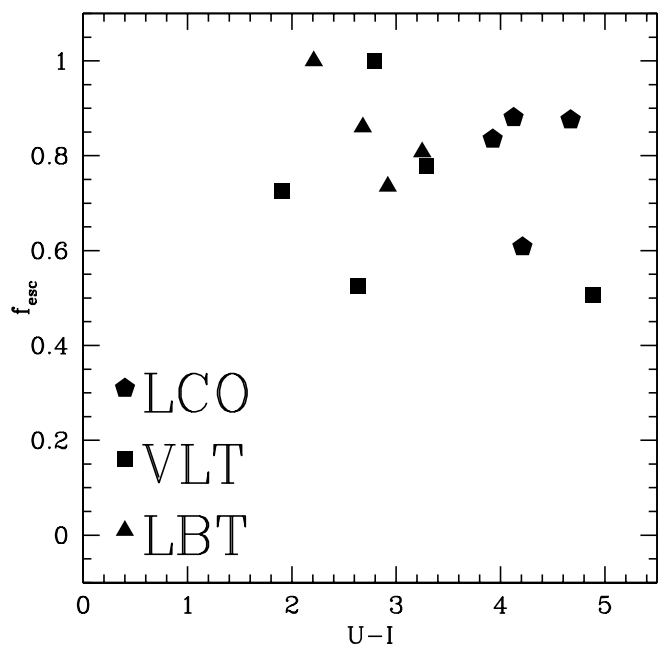}
\caption{The dependence of LyC escape fraction on the observed U-I color
for our faint AGNs. No obvious trend has been detected.
Three objects from Table \ref{sample} have no information on the
U-band magnitude and are not plotted here.
}
\label{fesccol}
\end{figure}

Fig.\ref{fesccol} shows the LyC escape fraction of our faint AGNs versus
the observed U-I color. No obvious trend is present, indicating that 
the typical color selection for $z\sim 4$ AGNs is not causing a notable
effect on the LyC transmission for our sample.

The fact that we do not find any change in the properties of the
escape fraction of faint AGNs ($L\sim L^*$) with respect to the brighter
QSO sample ($L\sim 10^3 L^*$) possibly indicates that
even fainter AGNs ($L\sim 10^{-2} L^*$) at $z\sim 4$ could
have an escape fraction larger than 75\% and possibly close to 100\%.
Although a gradual decrease of the escape fraction with decreasing
luminosity would be expected by AGN feedback models (see e.g. Menci et
al. 2008, Giallongo et al. 2012),
this trend becomes milder as the redshift increases especially at $z>4$.
For this reason, in the following we assume that AGNs brighter
than $L\sim 10^{-2} L^*$ ($M_{1450}\sim -18$) have $f_{esc}\sim 75\%$.
In a similar way, Cowie et al. (2009) and Stevans et al. (2014) show
values of $f_{esc}$ close to unity both for bright QSOs
and for much fainter Seyfert
galaxies at lower redshifts. We will use this assumption in the
next sections to derive the contribution of the faint AGN population to the
ionizing background at $z\sim 4$ and to make some speculations on the role
of accreting SMBHs to the reionization process at higher redshifts.

\subsection{The HI ionizing background at $z\sim 4$ produced by faint AGNs}

In order to evaluate the contribution of faint AGNs to the HI ionizing
background at $z\sim 4$ we assume that the LyC escape
fraction is at least
75\% for all the accreting SMBHs, from $M_{1450}=-28$ down to $M_{1450}=-18$,
which correspond approximately to a luminosity range
$10^{-2} L^*\le L\le 10^{2} L^*$, as carried out by Giallongo et al. (2015).

We use the same method adopted by Giallongo et al. (2015) to compute
the UVB (in units of $10^{-12}$ photons per second, $\Gamma_{-12}$) starting
from the AGN luminosity function and
assuming a mean free path of 37 proper Mpc at $z=4.0$, in agreement
with Worseck et al. (2014). At this aim, we adopt a spectral slope
$\alpha_\nu$ of -0.44 between 1200 and 1450 {\AA} rest frame and -1.57
below 1200 {\AA} rest frame, following Schirber \& Bullock
(2003). As discussed in Giallongo et al. (2015) and Cristiani et
al. (2016), this choice is almost equivalent to assume a shallower
slope of -1.41 below 1450 {\AA}, as found by Shull et al. (2012) and
Stevans et al. (2014).

In Table \ref{tabgamma} we compute the value of the HI ionizing UVB
$\Gamma_{-12}$ assuming $f_{esc}(LyC)=75\%$ and considering different
parameterizations of the $z=4$ AGN luminosity function according to the
different renditions found in the recent literature (Glikman et
al. 2011; Giallongo et al. 2015; Akiyama et al. 2017; Parsa et
al. 2018). When these luminosity functions
are centered in a different redshift bin,
we shift them to z=4 applying a density evolution of a factor of 3 per
unit redshift (i.e. $10^{-0.43z}$), according to the results provided
by Fan et al. (2001) for the density evolution of bright QSOs.
Here we assume that the bright and faint AGN populations evolve at the
same rate, which may not be completely true
(e.g. AGN downsizing, see Hasinger et al. 2005).

We provide the HI photo-ionization rate both at $M_{1450}\le -23$ and
at $M_{1450}\le -18$. The values in parenthesis show the fraction of
the UVB contributed by QSOs ad AGNs w.r.t. the value of
$\Gamma_{-12}=0.85$ found by Becker \& Bolton (2013, hereafter BB13) at z=4.
This value is required to keep the Universe ionized at these redshifts.
If we adopt a UVB of
$\Gamma_{-12}=0.55$ found by Faucher-Giguere et al. (2008, hereafter FG08)
at $z=4$, then the fractional values in Table \ref{tabgamma}
should be increased by a factor of 1.55.

\begin{table}
\caption{The HI photo-ionization rate $\Gamma_{-12}$ produced by AGN at
$z\sim 4$}
\label{tabgamma}
\centering
\begin{tabular}{l c c}
\hline
\hline
Luminosity Function & $\Gamma_{-12}$ &
$\Gamma_{-12}$ \\
 & $M_{1450}\le -23$ & $M_{1450}\le -18$ \\
\hline
Glikman et al. (2011) & 0.140 (16.5\%) & 0.307 (36.3\%) \\
Giallongo et al. (2015) & 0.208 (24.6\%) & 0.617 (72.9\%) \\
Akiyama et al. (2017) & 0.113 (13.4\%) & 0.135 (15.9\%) \\
Parsa et al. (2018) & 0.088 (10.4\%) & 0.255 (30.0\%) \\
\hline
\hline
\end{tabular}
\\
The LyC escape fraction $f_{esc}(LyC)$ is fixed to 75\% in the luminosity range
$-28\le M_{1450}\le -18$. In parenthesis we compute the fraction of the UVB
w.r.t. the value of $\Gamma_{-12}=0.85$ provided by BB13 at z=4.
The luminosity functions of Glikman et al. (2011) and
Akiyama et al. (2017) have magnitude limits which are significantly
brighter than the adopted integration limit $M_{1450}=-18$ and are extrapolated.
The  Giallongo et al. (2015) and Parsa et al. (2018) luminosity functions
instead provide an estimate of the AGN space density close to the adopted
integration limit of $M_{1450}=-18$.
\end{table}

If we consider the contribution of QSOs and AGNs down to $L\sim L^*$
($M_{1450}\le -23$) we find an emissivity which is always 10-25\% of the
HI photo-ionization rate provided by BB13, irrespectively of the adopted
parameterization of the luminosity function. This fraction rises to
15-38\% if the FG08 UVB is considered. Thus we can conclude, in agreement
with Cristiani et al. (2016), that bright QSOs can provide 10-40\%
of the whole ionizing UVB at $z\sim 4$. This conclusion is robust with
respect to the
adopted UVB and for different parameterizations of the AGN luminosity
function at these redshifts.

For fainter AGNs ($M_{1450}\le -18$) the situation depends critically
on the adopted luminosity function at $z\sim 4$ and on the exact value
of the HI photo-ionization rate measured through the Lyman-$\alpha$
forest statistics. Assuming an UVB by BB13, then the contribution of
faint AGNs is between 16 and 73\%, adopting respectively the
luminosity function of Akiyama et al. (2017) and Giallongo et
al. (2015). It is worth noting here that
there are still large uncertainties on the observed space density of
$L\sim L^*$ AGNs at $z\ge 4$ and that the luminosity function of
Akiyama et al. (2017) is systematically lower, by a factor of 10, than
the one of Glikman et al. (2011) at $M_{1450}\sim -23$ (see Fig. 18
of Akiyama et al. 2017).
While the result of Glikman et al. (2011) can be
considered as a lower limit for the space density of QSOs at
$z\sim 4$, since it has been derived from a spectroscopically complete
sample of point-like type 1 QSOs only, it could be affected by large
uncertainties related to the completeness corrections. For these reasons,
a firm value of the
luminosity function at $M_{1450}\sim -23$ is still required.

Recently, Parsa et al. (2018) have revised the results of Giallongo et
al. (2015), deriving a different parameterization for the luminosity
function at $z\ge 4$. It should be noted, however, that the space
density at $M_{1450}\ge -21$ in Parsa et al. (2018) at $z=4.25$ is
similar to the one by Giallongo et al. (2015). Their contribution to
the UVB is significantly lower than Giallongo et al. (2015) since they
adopt a rather flat luminosity function, with a normalization around
the break ($M_{1450}\sim -23$) which is significantly lower than the
space density provided both by Giallongo et al. (2015) and by Glikman
et al. (2011). As discussed extensively in Giallongo et al. (2012,
2015), the bulk of the ionizing UVB comes directly from objects close
to the break of the luminosity function, thus exactly where the Parsa
et al. (2018) fit is below the observed number density of AGNs by
Glikman et al. (2011). It
is thus not surprising that their luminosity function is providing
only 30\% of the HI photo-ionization rate at z=4.

\begin{figure}
\centering
\includegraphics[width=9cm,angle=0]{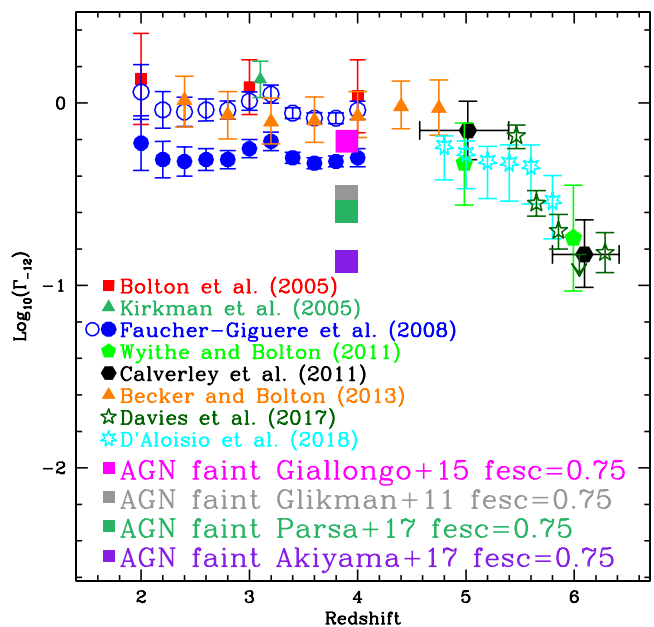}
\caption{The HI photo-ionization rate measured by different estimators from the
literature as a function of redshift. The magenta square shows the
estimated value for the emissivity of AGNs at $z\sim 4$, assuming a
luminosity function of Giallongo et al. (2015) down to $M_{1450}=-18$.
The grey square indicates the contribution of
faint AGNs adopting a luminosity function of Glikman et al. (2011).
The green and violet squares indicate the AGN emissivity assuming a
luminosity function of Parsa et al. (2018) and Akiyama et al. (2017),
respectively.
A LyC escape fraction of 75\% has been assumed. The UVB derived by
FG08 has been shown with filled blue circles, while we represent with open
blue circles the same data of FG08 rescaled with a different temperature-density
relation, as described in Calverley et al. (2011).
}
\label{gamma}
\end{figure}

Summarising, the faint AGN population at $z\sim 4$ are able to contribute,
adopting respectively the Glikman et al. (2011) and Giallongo et al. (2015)
luminosity function,
to 36-73\% of the UVB, assuming the BB13 determination, or 56-100\% of the
UVB, assuming FG08 measurement, as shown in Fig. \ref{gamma}.
If the luminosity functions of Akiyama et al. (2017) or Parsa et al. (2018)
are adopted, instead, the faint AGN population is able to
provide only 16-30\% of the BB13 UVB, or 25-47\% of the FG08 UVB, respectively.
Our conclusion is that, modulo the uncertainties on the UVB determination
and on the knowledge of the $z\sim 4$ AGN luminosity function,
the population of accreting super-massive black holes at
these redshifts can contribute to a non-negligible fraction of the
HI photo-ionization rate. They can provide at least 16\% of the UVB in the
most conservative case, adopting the Akiyama et al. (2017) or Parsa et al.
(2018) luminosity functions, but it can reach 100\%
if the space density of
$L\sim L^*$ AGNs is comparable to that found by Glikman et al.
(2011) or Giallongo et al. (2015).
This result is based on the extrapolation
that the LyC escape fraction of AGNs remains constant
at $\sim 75\%$ level down to $M_{1450}\sim -18$. This assumption is
not unreasonable, according to Fig. \ref{fescmabs}, and should be checked
with future observations.

This result is fundamental to understand whether
faint AGNs are the main drivers of the Reionization epoch at $z>6$, under
the reasonable assumptions that the LyC $f_{esc}$ remains roughly
constant from $z\sim 4$ to $z\ge 6$ and that the space density of
faint AGNs is of the order of the one found by Giallongo et al. (2015)
up to $z\sim 6$. This conclusion will hold if we assume that the
physical properties of AGNs do not vary dramatically from $z=4$ to
$z\ge 6$. Given the similarity in the optical-UV spectra of faint AGNs
in the local Universe (Stevans et al. 2014) and bright QSOs at $z\ge 4$
(Prochaska et al. 2009; Worseck et al. 2014; Cristiani et al. 2016),
this assumption can be considered safe.

In addition, we can use the estimation of the UV background by faint
AGNs at z=4 provided in Table \ref{tabgamma} to derive a reasonable
upper limit to the relative escape fraction of the star-forming galaxy
population at this redshift. Assuming for example the HI ionizing
background of BB13 and the AGN Luminosity Function of Glikman et
al. (2011), a relative escape fraction of 4.8\% for the galaxy
population\footnote{We assume here the $z=3.8$ galaxy luminosity by Bouwens
et al. (2015) integrated down to $M_{1500}=-13$ and $\xi_{ion}=10^{25.27}$.}
should be assumed in order to complement the AGN contribution to
$\Gamma_{-12}$. Alternatively, an escape fraction of $\sim 2\%$
for the galaxy population
is needed to complement the AGN contribution if a luminosity function
by Giallongo et al. (2015) is assumed for AGNs at $z\sim 4$, or if we
adopt the Glikman et al. (2011) luminosity function but considering
the UV background by FG08. Finally, if we consider the luminosity
function by Giallongo et al. (2015) and the UVB by FG08, then all the
HI ionizing background is produced by AGNs and the upper limit to the
escape fraction for galaxies is close to zero. If confirmed,
these upper limits we
have obtained at $z=4$ for star-forming galaxies, i.e. $f_{esc,rel}$ of
$0-5\%$, are comparable to the values found at lower redshifts
(e.g. Grazian et al. 2016, 2017), indicating a mild evolution in
redshift of the LyC escape fraction for star-forming galaxies at
$z\lesssim 4$. For comparison, Cristiani et al. (2016) find a slightly
higher value of 5.5-7.6\% for the LyC escape fraction of $z\ge 4$ galaxies.
If this trend is not changing significantly at higher
redshifts, it would be difficult for the galaxy population to reach
values of $f_{esc,rel}\sim 15\%$, required in order to reach reionization
at $z\sim 7$ with stellar radiation only (Madau 2017).
At high-z there are however indications that the LyC photon production
efficiency $\xi_{ion}$ could be larger than the estimates at $z\le 2$
(Bouwens et al. 2016), thus relaxing a bit the $f_{esc,rel}\sim 15\%$
requirement. Moreover, the star-forming galaxy luminosity function
is steepening at higher redshifts (e.g. Finkelstein et al. 2015)
further relaxing the constraint on the galactic LyC escape fraction.
Considering the large uncertainties on the faint-end slope and on the
possible cut-off (at $M_{1500}\sim -13$, see e.g. Livermore et al. 2017 and
Bouwens et al. 2017) of the luminosity function, an escape fraction
of $\sim 4-11\%$ (assuming $\xi_{ion}=10^{25.3}$) can still be sufficient
for star-forming galaxies to keep the Universe ionized at $z>5$ (Madau 2017).

%-----------------------------------------------------------------

\section{Discussions}

\subsection{Selection of targets}

The extension of our results to the whole AGN
population at $z\sim 4$ depends critically on the sample adopted to carry
out the LyC escape fraction measurement.
The ideal sample should be an unbiased sub-sample of all the AGNs (both
obscured and of unobscured), without any biases against or in favour of
strong LyC emission. Our starting sample is a mix of optically and
X-ray selected AGNs from SDSS3-BOSS (Dawson et al. 2013), from
NDWFS/DLS survey (Glikman et al. 2011), and from Chandra X-ray
observations in the COSMOS field (Marchesi et al. 2016; Civano et
al. 2016), selected only in redshift ($3.6<z<4.2$) and in I-band
apparent magnitude ($21<I<23.5$).

The SDSS has searched for high-z QSOs adopting optical color criteria
based on dropouts, and thus this selection is not biased in favour of
strong LyC emitters, but could be biased against them.
A large value for $f_{esc}(LyC)$ indeed would
reduce the drop-out color used to select high-z sources. As shown in
Prochaska et al. (2009), the selection of $z\le 3.6$ QSOs in the
SDSS-DR7 (Abazajian et al. 2009) is instead biased against blue
($u-g<1.5$) objects, i.e. they are prone to select sightlines with
strong Lyman limit absorptions, active nuclei with strong emission lines
or low escape fraction.

On the other hand, the X-ray selection is not biased towards obscured
AGNs, but in our sample only 4 objects (25\%) have been selected through this
criterion. We do not find any difference between the properties of X-ray
and optically selected AGNs in our sample.
In the future, the exploitation of a complete X-ray
selected sample of (both obscured and unobscured) faint AGNs will
be instrumental to check whether type 2 AGNs do not emit LyC photons
(as found by Cowie et al. 2009) and to put firmer
constraints on the global emissivity of the whole population of
accreting SMBHs at high-z.

In summary, our sample can be representative of
the whole AGN population at $L\sim L^*$, and is not biased against or
in favour of strong LyC emitters, as shown in Fig.\ref{colormag} and
\ref{fesccol}.

\subsection{The fraction of BAL AGNs}

We have excluded from our sample only the AGNs which show strong
absorption features in the Lyman-$\alpha$ line, a population called
BAL QSOs. The large column density ($N_{HI}\ge
10^{19} cm^{-2}$) associated to the strong absorbing systems could imply a
complete suppression of the flux in the LyC region.

Allen et al. (2011) and Paris et al. (2014) have estimated the BAL
fraction of bright QSOs in the SDSS to be around 10-14\%. In
principle this particular class of AGNs should be included in our
sample, in order to have an estimate of the ionizing contribution of
the whole AGN population, as done by Cristiani et al. (2016).
However, it is not established yet whether all the BAL population has
negligible emission of LyC photons. As an example, in the
lensed BAL QSO APM08279+5255 at z=3.9, significant emission has been
detected in the LyC region (see Fig. 1 of Saturni et al. 2016).

Moreover, McGraw et al. (2017) show the relevant examples of appearing
and disappearing broad absorptions in SDSS QSOs at $z\sim 2-5$ on the
timescales of 1-5 rest-frame years for 2-4\% of the observed BAL
sample. A possible explanation of the BAL variability over multi-year
time scales imply changes either in the gas ionisation level or in the
covering factor, supporting transverse-motion and/or ionization-change
scenarios to explain BAL variations. This indicates that it is a
relatively fast phenomenon, which is probably due to the small scale
environment of the accreting SMBH (few kpc), and it is probably not
affecting its isotropic emission of ionizing photons on cosmological
timescales related to the QSO lifetime or on cosmological scales (Mpc).

For these reasons, we decided not to correct the value of LyC
$f_{esc}$ found with our sample for the fraction of BAL QSOs present
in the global population of active galactic nuclei.
In the future, a systematic study of the LyC escape fraction and
of the fraction of BAL objects at different luminosities will
be important to assess the total contribution of the whole AGN population
to the ionizing UVB.

\subsection{Reliability of LyC $f_{esc}$ measurement}

The uncertainties related to the LyC escape fraction measurements
described above mainly depend on the signal to noise ratio of the
spectra and on the determination of the systemic redshift of the AGN,
which can be obtained through the position of the OI 1305 emission line.
In our case, only for two AGNs (COSMOS1311 and SDSS04) we have good
quality spectra which allow us to refine the spectroscopic redshifts.
For the other AGNs, we rely on the published spectroscopic redshifts.

The shift in redshift applied to COSMOS1311 and SDSS04 is not
affecting our LyC $f_{esc}$ estimate. To avoid mismatch for the
position of the LyC region, we decided to measure the escape fraction
relatively far from the LyC break, i.e. between 892 and 905 {\AA} rest
frame. For example, considering the $\Delta z\sim 0.019$ correction
for COSMOS1311, it will move the 905 {\AA} rest frame wavelength to
908 {\AA}. In the unlikely case that the adopted shift is not correct,
we are still estimating the escape fraction in the LyC region. In
general, since the corrections to the observed redshifts for our
sample are low, we are not biasing our estimates towards higher value
of LyC $f_{esc}$ for our AGNs.

The S/N ratio of our spectra in the LyC region varies between 11 and
121, when integrated between 892 and 905 {\AA} rest frame and along
the slit. It translates into an uncertainty on the measured LyC escape
fraction of few percents ($\sim 2-15\%$).
This shows that our measurements of the
ionizing emissivity of faint AGNs is robust against statistical and
systematic errors.

We assume here that the contribution of the intrinsic slope of the AGN
has negligible impact on the flux ratio between 900 and 930 {\AA} rest
frame, since this wavelength interval is relatively limited. Since the
typical spectral slope of AGNs is $f_\nu\propto \nu^{\alpha_\nu}$ with
$\alpha_\nu$ approximately between -0.5 and -1.0, it turns out that
the impact on the LyC escape fraction estimate is to decrease the
observed LyC $f_{esc}$ w.r.t. the true one. Correcting e.g. for the
intrinsic spectral slope of AGN, indeed, provides only a negligible
correction of the order of 2\%, assuming e.g. $\alpha_\nu=-0.7$.

We have avoided also the region within
the Stromgreen sphere of the AGN itself, the so-called proximity region.
If part of the proximity region is included in our calculations,
i.e. measuring
the ionizing radiation between 892 and 910 {\AA} rest frame instead of
limiting us to 905 {\AA}, we obtain an escape fraction which is on
average 4\% higher for our 16 AGNs, but with large scatter from object
to object. As can be seen from Fig.\ref{sdss04} and \ref{sdss36}, the
IGM absorption can vary a lot from different lines
of sight, and in some cases it turns out that the proximity region
between 905 and 910 {\AA} are affected by an intervening absorbers
(e.g. at $\lambda\sim 907$ {\AA} rest frame for SDSS36 in Fig.\ref{fesc}).

Indeed, if we correct our estimate of the escape fraction for the flux
decrement due to intervening absorbing systems between 930 and 900
{\AA} rest frame, adopting e.g. Inoue et al. (2014) at z=4 or
Prochaska et al. (2009) at z=3.9, we obtain a corrected LyC escape fraction
which is close to 100\%.
This indicates that our method provides a conservative
{\em lower} limit to the LyC escape fraction of faint AGNs, and there
are robust indications that the corrected estimate could be close to
$f_{esc}=100\%$ for our objects.

\subsection{The LyC escape fraction of low luminosity AGNs}

There are two different aspects of the problem on the sources of
ionizing photons: 1) are the LyC photons emitted by stars or are they the
result of accretion onto SMBHs ?; 2) are the ionizing photons
able to escape into the IGM as the result of
stellar feedback (winds, SNe), or as the result of AGN energetic feedback
into the ISM ?
We can try to derive here some hints. If the carving of
free channels in the ISM of a galaxy has been driven by an ubiquitous
mechanism such as supernovae or stellar winds, then we can expect that
a significant LyC escape fraction must be a common and widespread
phenomenon among SFGs. Since this is not the case (e.g. Grazian et
al. 2016, 2017; Japelj et al. 2017), and Lyman Continuum Emitters are
rare and peculiar cases both in the local Universe and at high-z
(e.g. Izotov et al. 2016a,b; De Barros et al. 2016, Vanzella et al. 2016;
Shapley et al. 2016, Bian et al. 2017), we can probably
exclude that the cleaning of free paths in the galaxies ISM is
due to SNe or stellar wind, but it could be caused by rarer phenomena
like accreting BHs, such as AGNs or X-ray binaries (XRBs).

At this point, it is important to explore the dependencies of the
LyC escape fraction on the luminosities of the AGNs.
Interestingly, Kaaret et al. (2017) find that the Lyman continuum
emitting galaxy Tol 1247-232 ($f_{esc,rel}=21.6\%$, Leitherer et al. 2016)
has been detected
in X-ray as a point-like source by Chandra, and it shows also X-ray
variability by a factor of 2 in few years, thus being probably a
low-luminosity AGN ($Lx\sim 10^{41} erg/s$) at $z=0.048$. Another LyC
source, Haro 11 with $f_{esc,abs}\sim 3\%$ (Leitet et al. 2011),
has been detected in X-ray
as a bright point source with a very hard spectrum (Prestwich et
al. 2015). In addition to these two galaxies, Borthakur et al. (2014)
show an example of a LyC emitter (J0921+4509, $f_{esc,abs}\sim 20\%$),
which has been detected in hard X-rays by XMM (Jia et al 2011),
possibly revealing its AGN nature.

We can link our observations at $z\sim 4$ with the results by Kaaret
et al. (2017) for extremely faint
AGNs in the local Universe. The LyC escape fraction
of AGNs with $Lx=10^{44} erg/s$ is substantial ($\sim 80-100\%$) at $0\le
z\le 4$ as shown by this work and by the recent literature (Cowie et al. 2009;
Stevans et al. 2014; Prochaska et
al. 2009; Cristiani et al. 2016), while it is only few percents for
the two faint AGNs of Kaaret
et al. (2017), with $Lx=10^{41} erg/s$ at $z\sim 0$.
As we will show in the following, this is not in contrast with our conclusions
above.

The total UV absolute magnitudes of Tol 1247-232 and Haro 11 are $\sim
-20/-21$, which translate into effective magnitudes of $\sim -15/-16$
at 900 {\AA} rest frame, given the observed values for their LyC
escape fraction. Starting from the X-ray luminosity of $Lx\sim 10^{41}
erg/s$ measured by Kaaret et al. (2017) for their central sources, and
assuming that the optical luminosities are scaling as $Lx\propto
L_{UV}^{0.6}$ (Lusso \& Risaliti 2016), we derive an absolute magnitude of
$M_{1450}\sim -12$ for the central AGNs in Tol 1247-232 and Haro 11.
This implies that the ionizing radiation by these two galaxies cannot
be entirely produced by the central engines, even assuming an escape
fraction of 100\% for the central AGN, but it is possibly emitted also by
the surrounding stars, once the AGN has cleaned the surrounding ISM.
As a consequence, a large mechanical power, probably available only in
accreting SMBHs, can drive the emission of copious amount of ionizing
photons, even in very low mass or low luminosity galaxies, as
suggested by theoretical models (Menci et al. 2008; Giallongo et al. 2012;
Dashyan et al. 2018) or found by
observations (e.g., Penny et al. 2017). If a linear relation between $Lx$
and $L_{opt}$ is instead adopted for these two AGNs, then their
magnitudes will be $M_{1450}\sim -15/-16$. In this case, it is
possible that the ionizing flux is entirely provided by the central
AGN, with an effective escape fraction $\sim 50-100\%$, indicating
that there is a relatively mild evolution of $f_{esc}$ with the AGN
luminosity over a very broad range of absolute magnitudes.

Moreover, it is worth pointing out that the two galaxies studied
by Kaaret et al. (2017) lie within the pure star-forming region of the
Baldwin-Phillips-Terlevich (BPT, \cite{bpt}) diagram and there are
no indications
for the presence of an active nucleus at other wavelengths (Leitet et
al. 2013). For this reason, these objects have not being considered
in the AGN population in the previous studies, which are thus
underestimating the correct space density of accreting SMBHs at faint
luminosities. More interestingly, these peculiar sources
are not taken into account when one investigates
the AGN contribution to the ionizing UVB. This point is
corroborated also by the results of Cimatti et al. (2013) and Talia et
al. (2017), who showed that there are AGNs at $z>2$ which are not showing any
kind of nuclear activity signature in their optical spectra. In the
local Universe, Chen et al. (2017) concluded that 30\% of the AGNs
observed by NuSTAR are not detected in soft X-ray, Optical or IR, thus
evading the typical AGN selection criteria. In the future, detailed
and complete census of the whole AGN population at high
redshifts and at faint luminosities will allow a
better understanding of the reionization process.

\subsection{The issue of too high IGM temperature with an AGN dominating
HeII reionization}

A scenario where HI reionization is driven mainly by AGNs has the draw
back of heating the IGM to too high temperature at the epoch of HeII
reionization, which is foreseen at $z\sim 4$ (Worseck et al. 2016).
Such issue is present in
models of AGN dominating both HI and HeII reionizations, as discussed
in D'Aloisio et al. (2017, 2018). Their simulations show that an
AGN-dominated model is in tension with the available constraints on
the thermal history of the intergalactic medium, since the early HeII
reionization at $z\sim 4$ is heating up the IGM at a temperature
$T\sim 2\times 10^4$ K, well above the Lyman-$\alpha$ forest
temperature of $T\sim 10^4$ K, inferred by Becker et al. (2011)
through the comparison with hydrodynamic simulations.

Recently, Puchwein et al. (2018) draw a similar conclusion, i.e. that
models with a large AGN contribution to the high-z UVB are
disfavored by the low IGM temperature determination of Becker et
al. (2011). It is worth pointing out, however, that measuring the IGM
temperature from the Lyman-$\alpha$ forest opacity is very
challenging. At present, the existing measurements of $T_{IGM}$ show
a large variance, with differences by even a factor of 2-2.5 between
different methods ($T_{IGM}\sim 10^4-2.5\cdot 10^4 K$ at $z\ge 3$, see
e.g. Puchwein et al. 2018; Hiss et al. 2017; Garzilli et al. 2012; Lidz et
al. 2010). Clearly, more observations and detailed simulations are
needed in order to understand the origin of these discrepancies. Given the
large uncertainties that are still present on the determination of the
IGM temperature at $z\ge 3$, a scenario where AGNs give a
significant contribution to the reionization cannot be ruled out.

Moreover, it is also possible that the extreme UV spectra of
low-luminosity AGNs are softer than that of bright QSOs, due to the
contribution of the host galaxy to the escaping LyC photons,
once the active SMBH has opened a clear
path into the ISM. As a consequence, the impact of low-luminosity AGNs
and their host galaxies to the HeII reionization could be in better
agreement with the predictions of D'Aloisio et al. (2017, 2018) and
Puchwein et al. (2018).

\subsection{Future activities}

In this paper we have shown that $z\sim 4$ AGNs with $L\gtrsim L^*$ have
in general large escape fraction of HI ionizing photons, and can contribute
to $\ge 50\%$ of the UVB at these redshifts.
With the present data, it is not obvious to conclude whether AGNs are
the main drivers of the HI reionization at $z>6$, since their number
densities are still uncertain at $z\ge 5$ (Giallongo et al. 2015; Ricci et
al. 2017; Parsa et al. 2018; Matsuoka et al. 2017; Onoue et al. 2017).

In order to give further insights into this important open question,
three major advances are required in the field of high-z AGNs:
1-it is crucial at this point to confirm
or disprove with large number statistics and at fainter
luminosities ($L\le L^*$) the result found in this paper, i.e. that
faint AGNs have substantial LyC escape fraction ($f_{esc}\sim 75\%$), similarly
to very bright high-z QSOs.
2-the luminosity function at $z\ge 4$ is still poorly sampled, especially
around $L\sim 0.1-1.0 L^*$, where the bulk of ionizing photons is expected.
3-the current measurements of the photo-ionization rate $\Gamma_{-12}$
at $z>2$ are still uncertain
(e.g. FG08 vs BB13), as discussed in the previous section.
These are three fundamental steps in order
to understand whether faint AGNs are the potential drivers of the
reionization process.
In this paper we have started to answer the questions summarised
in point 1, i.e. the escape fraction of faint AGNs.
In a future paper we will explore the dependencies of the LyC escape
fraction on the physical properties of our faint AGN sample at
$z\sim 4$. In Giallongo et
al. (in prep.) we are investigating the luminosity function of
AGNs at $z\ge 4$ at $M_{1450}\ge -21$. In order to give precise
answers to point 2, however, a dedicated survey of $L\sim 0.1-1.0 L^*$
AGNs (both of type 1 and type 2) is required.
The exact determination of the ionizing UVB has been discussed
extensively by many authors. It is not trivial to directly translate
the observations of the Lyman forest opacity of high-z QSOs into an estimate of
the UVB. For example, BB13 explore the
differences between their UVB determination and that of FG08.
They ascribe the discrepancy to the assumed
temperature-density relation (due primarily to their lower IGM
temperatures) and to the effect of
peculiar velocities and thermal broadening. In the future, detailed
simulations of the IGM at high spatial resolution will be probably able to
address these issues in more detail.

%-----------------------------------------------------------------

\section{Summary and Conclusions}

Most papers related to the hydrogen reionization at high-z
state that star-forming galaxies are the most obvious and natural
mechanism for producing the required ionizing photons.
An alternative scenario is possible, i.e., that faint AGNs could provide
a great fraction of the HI ionizing UVB, at least at $z\sim 4$.
This has important implications for the role of galaxies and AGNs to the
reionization of the Universe.

We selected 16 AGNs at $z\sim 4$ in a magnitude
range $-25.1\lesssim M_{1450}\lesssim -23.3$ (i.e. $L^*\lesssim L\lesssim
7L^*$) with the aim of measuring the LyC escape fraction of a
representative sample of relatively faint AGNs. We have shown that with typical
exposure times of $t_{exp}\sim 2-6$ hours per target at 6-8 meter class
telescopes, equipped with UV sensitive instruments (e.g. FORS2 and
MODS1-2), the quality of the acquired spectra is high enough to study the
LyC emission of $L\sim L^*$ AGNs. Our limited sample is already
suggesting a relatively large escape fraction of HI ionizing photons
($f_{esc}\ge 75\%$) for the whole AGN population with $L\gtrsim L^*$ at very high
confidence level ($S/N\sim 10-120$). Therefore, the ionizing
properties of the faint AGN population at
$z\sim 4$ are similar to those of the brightest QSOs, i.e.
$M_{1450}\sim -30$ or $L\sim 10^3 L^*$, at the same redshift.

Assuming the luminosity functions of Glikman et al. (2011) or
Giallongo et al. (2015) at $z\sim 4$, and extrapolating
the AGN contribution down to a magnitude of $M_{1450}\sim -18$, AGNs
can provide between 36 and 73\% of the UVB measured by BB13. If the
UVB by FG08 is considered instead, the integrated AGN contribution
rises up to 56-100\%. Adopting other luminosity functions (e.g. Parsa et
al. 2018; Akiyama et al. 2017) gives a lower contribution (16-30\% of
the ionizing UVB measured by BB13).

Based on these results, we conclude that faint ($L\sim L^*$) AGNs could provide
a crucial contribution to the cosmological UV background up to $z=4$,
and, if the large escape fraction and high space densities for AGNs
are confirmed also at $z\sim 5-7$, they could be the main responsibles
for the reionization of the Universe. This result is in agreement with
recent models which are showing that a large contribution from AGNs
to the ionizing background is sufficient to account for the observed
probability distribution function of the opacity $\tau_{GP}$ in the
lines of sight of bright $z\sim 6$ QSOs (Chardin et al. 2016,
2017). D'Aloisio et al. (2017) reach a similar conclusion, though
finding that an AGN driven reionization scenario heats the IGM to too
high temperature at the epoch of HeII reionization.
It is worth mentioning that the present measurements of the
IGM temperature at $z>3$ are characterized by a large variance,
due to the difficulties in comparing Lyman-$\alpha$ forest opacity
observations and simulations (e.g. Hiss et al. 2017; Puchwein et al. 2018).

In the future, it will be possible to substantiate these conclusions
by extending at fainter luminosities ($L<L^*$) and to the BAL class
the present analysis on the LyC escape fraction of AGNs, by measuring
with high accuracy the AGN space density near the break ($M_{1450}\sim
-23$) of the Luminosity Function at $z\ge 4$ and finally by estimating
in an unbiased and possibly direct way the ionizing UVB at $z\ge 4$
and the IGM temperature.

%-----------------------------------------------------------------

\begin{acknowledgements}
We warmly thank the anonymous referee for her/his useful suggestions and
constructive comments that help us to improve this paper.
AG and EG warmly thank Piero Madau and Enrico Garaldi for useful discussions.
The LBT is an international collaboration among institutions in the
United States, Italy, and Germany. LBT Corporation partners are The
University of Arizona on behalf of the Arizona university system;
Istituto Nazionale di Astrofisica, Italy; LBT
Beteiligungsgesellschaft, Germany, representing the Max-Planck
Society, the Astrophysical Institute Potsdam, and Heidelberg
University; The Ohio State University; and The Research Corporation,
on behalf of The University of Notre Dame, University of Minnesota,
and University of Virginia.
This paper used data obtained with the MODS spectrographs built with
funding from NSF grant AST-9987045 and the NSF Telescope System
Instrumentation Program (TSIP), with additional funds from the Ohio
Board of Regents and the Ohio State University Office of Research.
Based on observations collected at the European Organisation for Astronomical
Research in the Southern Hemisphere under ESO programme 098.A-0862.
\end{acknowledgements}


\end{document}